\documentclass[letterpaper]{article}
\usepackage{setspace}
\usepackage{amsthm,amsmath,amssymb,natbib}
\RequirePackage[colorlinks,citecolor=blue,urlcolor=blue]{hyperref}

\usepackage{xspace,soul}
\usepackage{graphicx}

\newcommand{\Ex}{\mathbb{E}}

\newcommand{\R}{\textsf{R}\xspace}

\def\balpha{\pmb{\alpha}}
\def\bbeta{\pmb{\beta}}
\def\bgamma{\pmb{\gamma}}
\def\btheta{\pmb{\theta}}
\def\bphi{\pmb{\phi}}
\def\bpsi{\pmb{\psi}}
\def\bB{\pmb{B}}
\def\bD{\pmb{D}}
\def\bH{\pmb{H}}
\def\bS{\pmb{S}}
\def\bX{\pmb{X}}

\begin{document}

\title{openWAR: An Open Source System for Evaluating Overall Player Performance in Major League Baseball}
\author{
Benjamin S. Baumer \\
%Department of Mathematics \& Statistics \\
Smith College \\
%Northampton, MA 01063, USA \\
\texttt{bbaumer@smith.edu}
\and
Shane T. Jensen\\
%Department of Statistics \\ 
The Wharton School\\University of Pennsylvania\\
\texttt{stjensen@wharton.upenn.edu}
\and
Gregory J. Matthews\\
%Department of Biostatistics and Epidemiology\\
Loyola University Chicago\\
\texttt{gjm112@gmail.com}
}

\maketitle

\begin{abstract}
Within sports analytics, there is substantial interest in comprehensive statistics intended to capture overall player performance.  In baseball, one such measure is Wins Above Replacement (WAR), which aggregates the contributions of a player in each facet of the game: hitting, pitching, baserunning, and fielding.  However, current versions of WAR depend upon proprietary data, ad hoc methodology, and opaque calculations.  We propose a competitive aggregate measure, $openWAR$, that is based on public data, a methodology with greater rigor and transparency, and a principled standard for the nebulous concept of a ``replacement" player.  Finally, we use simulation-based techniques to provide interval estimates for our $openWAR$ measure that are easily portable to other domains. 

Keywords: baseball, statistical modeling, simulation, \R, reproducibility %bootstrapping?
\end{abstract}

%%%%%%%%%%%%%%%%%%%%%%%%%%%%%%%%%%%%%%%%%%%%%%%%%%%%%%%%%%%%%%%%%%%%%%%

\section{Introduction}

In sports analytics, researchers apply statistical methods to game data in order to estimate key quantities of interest. In team sports, arguably the most fundamental challenge is to quantify the contributions of individual players towards the collective performance of their team.  In all sports the ultimate goal is winning and so the ultimate measure of player performance is that player's overall contribution to the number of games that his team wins.   Although we focus on a particular measure of player contribution, wins above replacement (WAR) in major league baseball, the issues and approaches examined in this paper apply more generally to any endeavor to provide a {\it comprehensive} measure of individual player performance in sports. 

A common comprehensive strategy used in sports such as basketball, hockey, and soccer is the \emph{plus/minus} measure \citep{kubatko2007starting,macdonald2011regression}.   Although many variations of plus/minus exist, the basic idea is to tabulate changes in team score during each player's appearance on the court, ice, or pitch.    If a player's team scores more often than their opponents while he is playing, then that player is considered to have a positive contribution. Whether those contributions are primarily offensive or defensive is not delineated, since the fluid nature of these sports make it extremely difficult to separate player performance into specific aspects of gameplay.  

In contrast, baseball is a sport where the majority of the action is discrete and player roles are more clearly defined. This has led to a historical focus on separate measures for each aspect of the game: hitting, baserunning, pitching and fielding.    For measuring hitting, the three most-often cited measures are batting average ($BA$), on-base percentage ($OBP$) and slugging percentage ($SLG$) which comprise the conventional ``triple-slash line" ($BA/OBP/SLG$).   More advanced measures of hitting include runs created~\citep{james1986bjh}, and linear weights-based metrics like weighted on-base average ($wOBA$) ~\citep{tango2007bpp} and extrapolated runs~\citep{furtado1999xr}.  Similar linear weights-based metrics are employed in the evaluation of baserunning~\citep{ubr}.

Classical measures of pitching include walks and hits per innings pitched ($WHIP$) and earned run average ($ERA$).  \cite{mccracken01pd} introduced defense independent pitching measures (DIPS) under the theory that pitchers do not exert control over the rate of hits on balls put into play.   Additional advancements for evaluating pitching include fielding independent pitching ($FIP$)~\citep{tangofip} and $xFIP$~\citep{studemanxfip}.  Measures for fielding include ultimate zone rating ($UZR$)~\citep{uzr}, defensive runs saved ($DRS$)~\citep{drs}, and spatial aggregate fielding evaluation ($SAFE$)~\citep{jensen2009bayesball}.  For a more thorough review of the measures for different aspects of player performance in baseball, we refer to the reader to~\cite{thorn1984hgb,lewis2004maw,albert2003cbb,schwarz2005ngb,tango2007bpp,baumer2013tsr}. 

Having separate measures for the different aspects of baseball has the benefit of isolating different aspects of player ability.  However, there is also a strong desire for a comprehensive measure of overall player performance, especially if that measure is closely connected to the success of the team.   The ideal measure of player performance is each player's contribution to the number of games that his team wins.  The fundamental question is how to apportion this total number of wins to each player, given the wide variation in the performance and roles among players.    

Win Shares~\citep{james2002win} was an early attempt to measure player contributions on the scale of wins.  Value Over Replacement Player \citep{vorp} measures player contribution on the scale of runs relative to a {\it baseline} player.  An intuitive choice of this baseline comparison is a ``league average" player but since average players themselves are quite valuable, it is not reasonable to assume that a team would have the ability to replace the player being evaluated with another player of league average quality.  Rather, the team will likely be forced to replace him with a minor league player who is considerably less productive than the average major league player.  Thus, a more reasonable choice for this baseline comparison is to define a ``replacement" level player as the typical player that is readily accessible in the absence of the player being evaluated.   

The desire for a comprehensive summary of an individual baseball player's contribution on the scale of team wins, relative to a replacement level player, has culminated in the popular measure of Wins Above Replacement (WAR).   The three most popular existing implementations of WAR are: $fWAR$ \citep{fangraphs}, $rWAR$ (sometimes called $bWAR$) \citep{bbref,rwar}, and $WARP$ \citep{warp}. A thorough comparison of the differences in their methodologies is presented in our supplementary materials. 

WAR has two virtues that have fueled its recent popularity. First, having an accurate assessment of each player's contribution allows team management to value each player appropriately, both for the purposes of salary and as a trading chip. Second, the units and scale are easy to understand. To say that Miguel Cabrera is worth about seven wins above replacement means that losing Cabrera to injury should cause his team to drop about seven games in the standings over the course of a full season. Unlike many baseball measures, no advanced statistical knowledge is required to understand this statement about Miguel Cabrera's performance. Accordingly, WAR is now cited in mainstream media outlets like \textit{ESPN}, \textit{Sports Illustrated},  \textit{The New York Times}, and the \textit{Wall Street Journal}.   

In recent years, this concept has generated significant interest among baseball statisticians, writers, and fans ~\citep{schoenfield2012war}.  WAR values have been used as quantitative evidence to buttress arguments for decisions upon which millions of dollars will change hands~\citep{rosenberg2012war}.  Recently, WAR has achieved two additional hallmarks of mainstream acceptance: 1) the 2012 American League MVP debate seemed to hinge upon a disagreement about the value of WAR~\citep{rosenberg2012war}; and 2) it was announced that the Topps baseball card company will include WAR on the back of their next card set~\citep{topps}. Testifying to the static nature of baseball card statistics, WAR is only the second statistic (after OPS) to be added by Topps since 1981.

\subsection{Problems with WAR}

While WAR is comprehensive and easily-interpretable as described above, the use of WAR as a statistical measure of player performance has two fundamental problems: a lack of uncertainty estimation and a lack of reproducibility.   Although we focus on WAR in particular, these two problems are prevalent for many measures for player performance in sports as well as statistical estimators in other fields of interest. 

WAR is usually misrepresented in the media as a known quantity without any evaluation of the uncertainty in its value.   While it was reported in the media that Miguel Cabrera's WAR was 6.9 in 2012, it would be more accurate to say that his WAR was {\it estimated to be} 6.9 in 2012, since WAR has no single definition.   The existing WAR implementations mentioned above ($fWAR$, $rWAR$ and $WARP$) do not publish uncertainty estimates for their WAR values.  As Nate Silver articulated in this 2013 ASA presidential address, journalists struggle to correctly interpret probability, but it is the duty of statisticians to communciate uncertainty~\citep{rickert2013}.  

Even more important than the lack of uncertainty estimates is the lack of reproducibility in current WAR implementations ($fWAR$, $rWAR$ and $WARP$).   The notion of reproducible research began with Knuth's introduction of \emph{literate programming}~\citep{knuth1984literate}.  The term \emph{reproducible research} first appeared about a decade later~\citep{claerbout1994hypertext}, but quickly attracted attention. \cite{buckheit1995wavelab} asserted that a scientific publication in a computing field represented only an advertisement for the scholarly work -- not the work itself. Rather, ``the actual scholarship is the complete software development environment and complete set of instructions which generated the figures"~\citep{buckheit1995wavelab}. Thus, the burden of proof for reproducibility is on the scholar, and the publication of computer code is a necessary, but not sufficient condition. Advancements in computing like the \texttt{knitr} package for \R~\citep{xie2014knitr} made reproducible research relatively painless. It is in this spirit that we understand ``reproducibility."

Interest in reproducible research has exploded in recent years, amid an increasing realization that many scientific findings are not reproducible~\citep{ioannidis2013,naik2011,zimmer2012,johnson2014,editorial2013,Economist2013}. Transparency in sports analytics is more tenuous than other scientific fields since much of the cutting edge research is being conducted by proprietary businesses or organizations that are not interested in sharing their results with the public.  

To the best of our knowledge, no open-source implementations  of $rWAR$, $fWAR$, or $WARP$ exist in the public domain and the existing implementations do not meet the standard for reproducibility outlined above. 	Two of the three methods use proprietary data sources, and the third implementation, despite making overtures toward openness, is still not reproducible without needing extra proprietary details about their methods.  This is frustrating since these WAR implementations are essentially ``black boxes" containing \emph{ad hoc} adjustments and lacking in a unified methodology~\footnote{For example, $rWAR$ and $fWAR$ are constrained to sum to 1000 in a season for no apparent substantive reason. See Section \ref{sec:repl} for a fuller discussion.}.   

\subsection{Contributions of openWAR}
	
We address both the lack of uncertainty estimates and the lack of reproducibility in Wins Above Replacement by presenting a fully transparent statistical model based on our {\it conservation of runs} framework with uncertainty in our model-based WAR values estimated by resampling methods.   In this paper we present $openWAR$, a reproducible and fully open-source reference implementation for estimating the Wins Above Replacement for each player in major league baseball.   

In Section~\ref{openwarmodel}, we introduce the notion of \emph{conservation of runs}, which forms the backbone of our WAR calculations. The central concept of our model is that the positive and negative consequences of all runs scored in the game of baseball must be allocated across four types of baseball performance: 1) batting; 2) baserunning; 3) fielding; and 4) pitching. While there are four components of $openWAR$, each is viewed as a component of our unified conservation of runs model.  

In contrast, the four components of WAR are estimated separately in each previous WAR implementation ($rWAR$, $fWAR$, or $WARP$) and these implementations only provide point estimates of WAR.  We employ resampling techniques to derive uncertainty estimates for $openWAR$, and report those alongside our point estimates. While the apportionment scheme that we outline here is specific to baseball, the resampling-based uncertainty measures presented in Section \ref{sec:interval} are generalizable to any sport. 

Our goal in this effort is to provide a coherent and principled fully open-source estimate of player performance in baseball that may serve as a reference implementation for the statistical community and the media.  Our hope is that in time, we can solidify WAR's important role in baseball by rallying the community around an open implementation.   In addition to the full model specification provided in this paper, our claim of reproducibility is supported by the simultaneous release of a software package for the open-source statistical computing environment \textsf{R}, which contains all of the code necessary to download the data and compute $openWAR$. 

\subsection{OpenWAR vs. previous WAR implementations}

In our approach, WAR for a player is defined as the sum of all of their marginal contributions in each of the four aspects of the game, relative to a hypothetical replacement level player after controlling for potential confounders (e.g. ballpark, handedness, position, etc.).    Previous WAR estimates, such as $rWAR, fWAR$, and $WARP$, serve as an inspiration for our approach but we make several key assumptions that differentiate our WAR philosophy from these previous efforts.  In addition to using higher resolution ball-in-play data than previous methods, we also have several differences in perspective. 

First, $openWAR$ is a retrospective measure of player performance -- it is not a measure of player ability to be used for forecasting. It is not context-independent, because we feel that context is important for accurate accounting of what actually happened.  Second, we control for defensive position in both our batting and fielding estimates. We do this at the plate appearance level, which allows for more refined comparisons of players to their appropriate peer group. Third, we believe that credit or blame for hits on balls in play should be shared between the pitcher and fielders.  We use the location of the batted ball to inform the extent to which they should be shared.    Fourth, we propose a new definition of replacement level based on distribution of performance beyond the 750 active major league players that play each day, which is different from existing implementations. Thus, $openWAR$ is not an attempt to reverse-engineer any of the existing implementations of WAR. Rather, it is a new, fully open-source attempt to estimate player performance on the scale of wins above replacement.

\section{Preliminaries: Expected Runs}\label{prelim}

A major hurdle in producing a reproducible version of WAR is the data source. $openWAR$ uses data published by Major League Baseball Advanced Media for use in their GameDay web application~\citep{mlbam}. A thorough description of the MLBAM data set obtainable using the $openWAR$ package is presented in our supplementary materials. 

Our $openWAR$ implementation is based upon a {\it conservation of runs} framework, which tracks the changes in the number of expected runs scored and actual runs scored resulting from each in-game hitting event.   The starting point for these calculations is establishing the number of runs that would be expected to score as a function of the current state of the game. Here, we illustrate that the \emph{expected run matrix}---a common sabermetric construction dating back to the work of~\cite{lindsey1959sdu,lindsey1961psd}---can be used to model th1ese quantities.~\footnote{The expected run matrix is also the basis for Markov Chain models, which have been used to, among other things, optimize batting order~\citep{freeze1974abb,pankin1978eop,bukiet1997mca,sokol2003rhb}.}

\label{deltas}
There are 24 different states in which a baseball game can be at the beginning of a plate appearance: 3 states corresponding to the number of outs (0, 1, or 2) and 8 states corresponding to the base configuration (bases empty, man on first, man on second, man on third, man on first and second, man on first and third, man of second and third, bases loaded).  A 25th state occurs when three outs are achieved by the defensive team and the half-inning ends.  

We define expected runs at the start of a plate appearance given the current state of an inning, 
$$
	\rho(o,b) = \Ex \, [\, R \, | \, startOuts=o, startBases=b \,] \, ,
$$
where $R$ is a random variable counting the number of runs that will be scored from the current plate appearance to the end of the half-inning when three outs are achieved. $startOuts$ is the number of outs at the beginning of the plate appearance, and $startBases$ is the base configuration at the beginning of the plate appearance.   The value of $\rho(o,b)$ is estimated as the empirical average of the number of runs scored (until the end of the half-inning) whenever a game was in state $(o,b)$.   Note that the value of the three out state is defined to be zero (i.e. $\rho(3,0) \equiv 0$).  

We can then define the {\it change} in expected runs due to a particular plate appearance as
$$
	\Delta \rho = \rho_{endState} - \rho_{startState} \,,
$$
where $\rho_{startState}$ and $\rho_{endState}$ are the values of the expected runs in the state at the beginning of the plate appearance and the state at the end of the plate appearance, respectively.    However, we must also account for the actual number of runs scored $r$ in that plate appearance, which gives us 
$$
\delta = \Delta \rho + r \,.
$$ 
For each plate appearance $i$, we can calculate $\delta_i$ from the observed start and end states for that plate appearance as well as the observed number of runs scored.   This quantity $\delta_i$ can be interpreted as the total run value that the particular plate appearance $i$ is worth.  Sabermetricians often refer to this quantity as RE24~\citep{re24}\footnote{RE for ``run expectancy" and 24 for the 24 distinct states}.

\section{openWAR Model} \label{openwarmodel}

%\subsection{Conservation of Runs Framework} 

The central idea of our approach to valuing individual player contributions is the assumption that every run value $\delta_i$ gained by the offense as a result of a plate appearance $i$ is accompanied by a corresponding $-\delta_i$ gained by the defense.   We call this principle our \emph{conservation of runs} framework.    The remainder of this section will outline a principled methodology for apportioning $\delta_i$ among the offensive players and apportioning $-\delta_i$ among the defensive players involved in plate appearance $i$.  
%Here, we make the assumption that on each play where the offense achieves $\delta$ value from the beginning to the end of a plate appearance, the defense has achieved $-\delta$ in allowing the play to occur.  

%GREG Comment: This looks strange because our data has a hat on it.   
%SHANE followup: I agree with Greg.  I removed the hat from delta.  It isn't like we are consistent about putting hats on all estimated 
% values anyways

\subsection{Adjusting Offensive Run Values} \label{adjustingoffensive}

As outlined above, $\delta_i$ is the run value for the offensive team as a result of plate appearance $i$.   We begin our modeling of offensive run value by adjusting $\delta_i$ for several factors beyond the control of the hitter or baserunners that make it difficult to compare run values across contexts. Specifically,  we want to first adjust for the ballpark of the event and any platoon advantage the batter may have over the pitcher (i.e. a left-handed batter against a right-handed pitcher).   We control for these factors by fitting a linear regression model to the offensive run values, 
\begin{eqnarray}
	\delta_{i} \,\, = \,\, \bB_i \,\cdot \, \balpha  \,\, + \,\, \epsilon_i \,,  \label{reg1}
\end{eqnarray}
where the covariate vector $\bB_i$ contains a set of indicator variables for the specific ballpark for plate appearance $i$ and an indicator variable for whether or not the batter has a platoon advantage over the pitcher.  The coefficient vector $\balpha$ contains the effects of each ballpark and the effect of a platoon advantage on the offensive run values.  Regression-based ballpark factors have been previously estimated by~\cite{acharya2008improving}. Estimated coefficients $\widehat{\balpha}$ are calculated by ordinary least squares using every plate appearance in our dataset.

The estimated residuals from the regression model (\ref{reg1}), 
\begin{eqnarray}
\hat{\epsilon}_i \,\, = \,\, \delta_i -  \bB_i \,\cdot \, \widehat{\balpha}
\end{eqnarray}
represent the portion of the offensive run value $\delta_i$ that is not attributable to the ballpark or platoon advantage, and so we refer to them as {\it adjusted} offensive run values.

\subsection{Baserunner Run Values} \label{baserunner}

The next task is determining the portion of $\hat{\epsilon}_i$ that is attributable to the baserunners for each plate appearance $i$ based on the following principle: {\it baserunners should only get credit for advancement beyond what would be expected given their starting locations,  the number of outs, and the hitting event that occurred}.     We can estimate this expected baserunner advancement by fitting a second regression model to our adjusted offensive run values, 
\begin{eqnarray}
\hat{\epsilon}_i \,\, = \,\, \bS_i \,\cdot \, \bbeta \,\, + \,\, \eta_i  \,, \label{reg2}
\end{eqnarray}
where the covariate vector $\bS_i$ consists of: 1) a set of indicator variables that indicate the specific game state (number of outs, locations of baserunners) at the start of plate appearance $i$ and; 2) the hitting event (e.g. single, double, etc.) that occurred during plate appearance $i$.   The 31 event types in the MLBAM data set that describe the outcome of a plate appearance are listed in our supplementary materials.   Estimated coefficients $\widehat{\bbeta}$ are calculated by ordinary least squares using every plate appearance in our dataset.   The estimated residuals from the regression model (\ref{reg2}), 
\begin{eqnarray}
\hat{\eta}_i \,\, = \,\,  \hat{\epsilon}_i -  \bS_i \,\cdot \, \widehat{\bbeta} \, , \label{reg3}
\end{eqnarray}
represent the portion of the adjusted offensive run value that is attributable to the baserunners.    If the baserunners take extra bases beyond what is expected, then $\hat{\eta}_i$ will be positive, whereas if they take fewer bases or get thrown out then $\hat{\eta}_i$ will be negative.   Note that $\hat{\eta}_i$ also contains the baserunning contribution of the hitter for plate appearance $i$.

We apportion baserunner run value, $\hat{\eta}_i$ amongst the individual baserunners involved in plate appearance $i$ based upon their expected base advancement compared to their actual base advancement.   If we denote $k_{ij}$ as the number of bases advanced by the $j^{th}$ baserunner after hitting event $m_i$, then we can use all plate appearances in our dataset to calculate the empirical probability
$$
	\hat{\kappa}_{ij} = \widehat{\Pr}( K \leq k_{ij} | m_i )
$$ 
that a typical baserunner would have advanced at least the $k_{ij}$ bases that baserunner $j$ did 
advance in plate appearance $i$.     If baserunner $j$ does worse than expected (e.g. not advancing from second on a single) then $\hat{\kappa}_{ij}$ will be small whereas if baserunner $j$ takes an extra base (e.g. scoring from second on a single), then $\hat{\kappa}_{ij}$ will be large.   These advancement probabilities $\hat{\kappa}_{ij}$ are used as weights for apportioning the baserunner run value, $\hat{\eta}_i$, to each individual baserunner, 
\begin{eqnarray}
	{\rm RAA}^{br}_{ij} = \frac{\hat{\kappa}_{ij}}{\sum_{j} \hat{\kappa}_{ij}} \cdot \hat{\eta}_i 
\end{eqnarray}
The value ${\rm RAA}^{br}_{ij}$ is the runs above average attributable to the $j^{th}$ baserunner on the $i^{th}$ plate appearance.

\subsection{Hitter Run Values}\label{hitter}

As calculated in (\ref{reg3}) above, $\hat{\eta}_i$ represents the portion of the adjusted offensive run value $\hat{\epsilon}_i$, that is attributable to the baserunners during plate appearance $i$.    The remaining portion of the adjusted offensive run value, 
\begin{eqnarray}
\hat{\mu}_i \,\, = \,\,  \hat{\epsilon}_i -   \hat{\eta}_i  \label{reg4}
\end{eqnarray}
is the adjusted offensive run value attributable to the hitter during plate appearance $i$.   Our remaining task for hitters is to calibrate their hitting performance relative to the expected hitting performance based on all players at the same fielding position.~\footnote{This is necessary because players who play more difficult fielding positions tend to be weaker hitters. In the extreme case, pitchers as a group are far worse hitters than those who play any other position. Thus, to evaluate the batting performance of pitchers without correcting for their defensive position would result in almost every pitcher being assigned a huge negative value for their batting performance. This would result in a dramatic undervaluation of pitchers (in the National League, at least) since they are obligated to hit while they are pitching.} We fit another linear regression model to adjust the hitter run value by the hitter's fielding position,
\begin{eqnarray}
\hat{\mu}_i \,\, = \,\, \bH_i \,\cdot \, \bgamma \,\, + \,\, \nu_i   \label{reg5}
\end{eqnarray}
where the covariate vector $\bH_i$ consists of a set of indicator variables for the fielding position of the hitter in plate appearance $i$.  Note that pinch-hitter (PH) and designated hitter (DH) are also valid values for hitter position.    Estimated coefficients $\widehat{\bgamma}$ are calculated by ordinary least squares using every plate appearance in our dataset.  The estimated residuals from this regression model, 
\begin{eqnarray}
	{\rm RAA}^{hit}_{i} = \hat{\nu}_i = \hat{\mu}_i - \bH_i \,\cdot \, \bgamma
\end{eqnarray}
represent the run values (above the average for the hitter's position) for the hitter in each plate appearance $i$.  

\subsection{Apportioning Defensive Run Values}\label{defensive}

As we discussed in Section~\ref{deltas}, each plate appearance $i$ is associated with a particular run value $\delta_i$, and we apportioned the offensive run value $\delta_i$ between the hitters and various baserunners in Sections~\ref{adjustingoffensive}-\ref{hitter}.   Now, we must apportion the defensive run value $-\delta_i$ between the pitcher and various fielders involved in plate appearance $i$.    

The degree to which the pitcher (versus the fielders) is responsible for the run value of a ball in play depends on how difficult that batted ball was to field.  Surely, if the pitcher allows a batter to hit a home run, the fielders share no responsibility for that run value. Conversely, if a routine groundball is muffed by the fielder, the pitcher should bear very little responsibility.   

We assign the entire defensive run value $-\delta_i$ to the pitcher for any plate appearance that does not result in a ball in play (e.g. strikeout, home run, hit by pitch, etc.).  For balls hit into play, we must estimate the probability $p$ that each ball-in-play would result in an out given the location that ball in play was hit.   

The MLBAM data set contains $(x,y)$-coordinates that give the location of each batted ball, and we use a two-dimensional kernel density smoother \cite[]{wand1994kernSmooth} to estimate the probability of an out at each coordinate in the field,
$$
	\hat{p}_i = f(x_i, y_i)
$$

\begin{figure}
	\centering
    \includegraphics[width=0.8\textwidth]{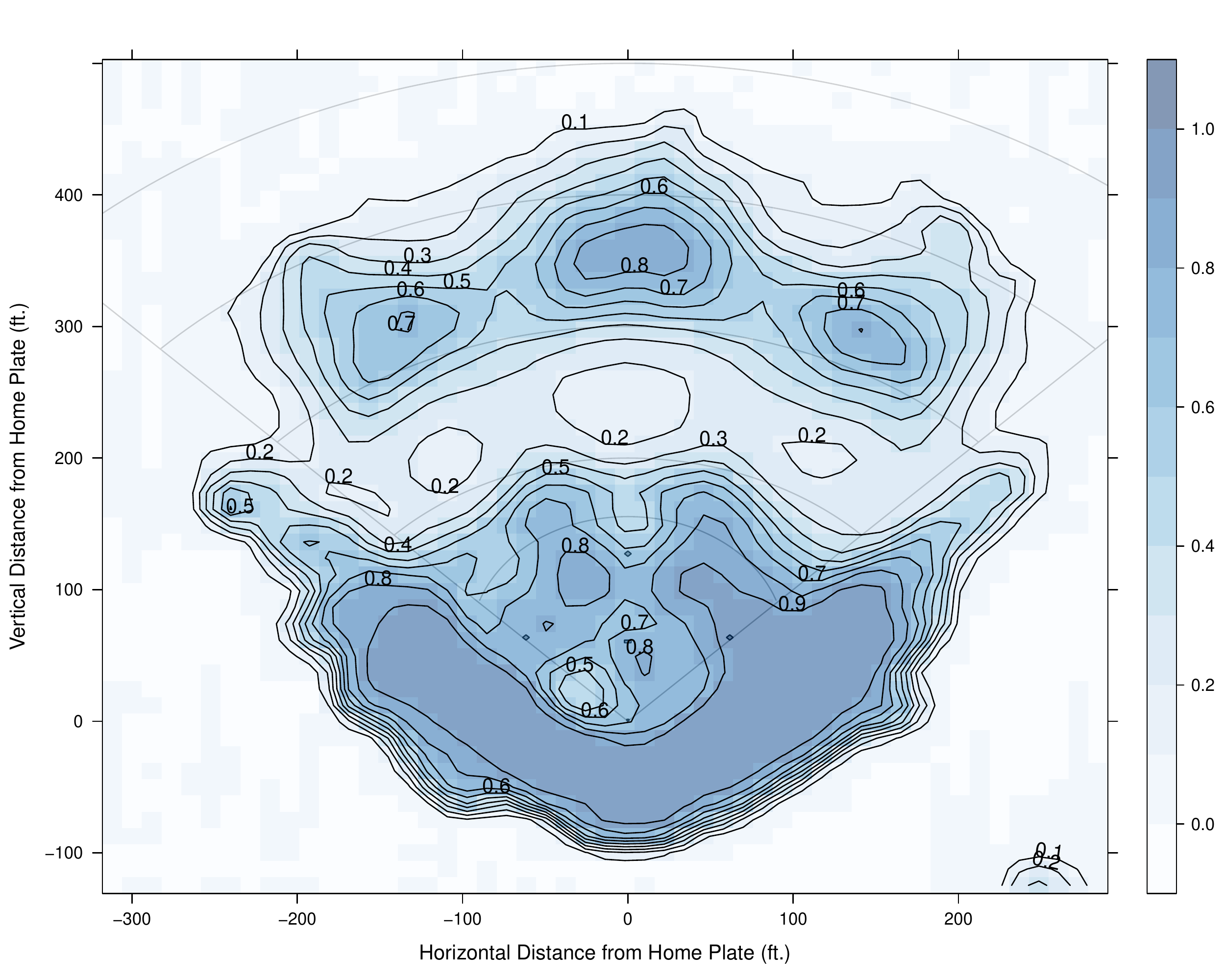}
    \caption{Contour plot of our estimated probability of an out $\hat{p}_i$ for a ball in play $i$ as a function of the coordinates $(x_i, y_i)$ for that ball in play.  Numerical labels give the estimated probability of an out at that contour line. }
    \label{fig:fielding}
\end{figure}

Figure \ref{fig:fielding} gives the contour plot of our estimated probability of an out, $\hat{p}_i$, for a ball in play $i$ hit to coordinate $(x_i, y_i)$ in the field.   For that ball in play $i$, we use $\hat{p}_i$ to divide the responsibility between the pitcher and the fielders.  Specifically, we apportion 

\begin{center}
\begin{tabular}{llll} 
$\delta^p_i$ & $=$ & $-\delta_i \cdot (1-p_i)$ &\qquad \text{to the pitcher}  \\
$\delta^f_i$ & $=$ & $-\delta_i \cdot p_i$ & \qquad \text{to the fielders} 
\end{tabular}
\end{center}

The fielders bear more responsibility for a ball in play that is relatively easy to field ($\hat{p}_i$ near 1) whereas a pitcher bears more responsibility for a ball in play that is relatively hard to field ($\hat{p}_i$ near 0). 

\subsection{Fielding Run Values}\label{fielding}

\begin{figure}
	\centering
  \includegraphics[width=0.8\textwidth]{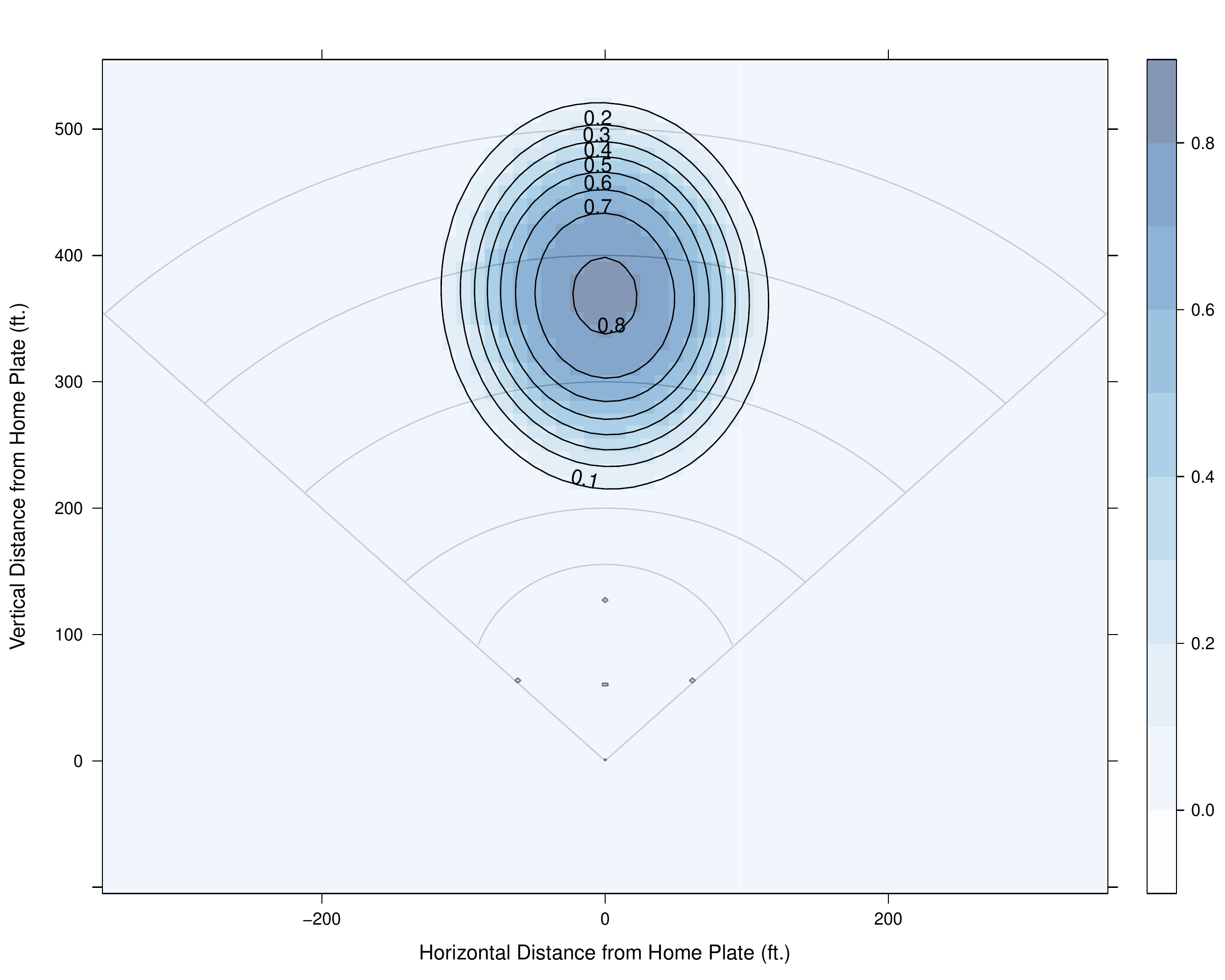}
  \caption{Contour plot of fielding model for centerfielders. The contours indicate the expected probability that any given centerfielder will catch a fly ball hit to the corresponding location on the field.}
  \label{fig:cf}
\end{figure}

In Section~\ref{defensive} above, we allocated the run value $\delta^f_i$ to the fielders.   We must now divide that run value amongst the nine fielders who are potentially responsible for ball in play $i$.    For each fielding position $\ell$, we use all balls in play to fit a logistic regression model, 
$$
	logit(p_{i \ell}) = \bX_i \, \cdot \, \btheta_\ell 
$$
where $p_{i \ell}$ is the probability that fielder $\ell$ makes an out~\footnote{Here we interpret ``making an out" as successfully converting a ball in play into at least one out.} on ball in play $i$ hit to coordinate ($x_i,y_i$) in the field.  The covariate vector $\bX_i$ consists of linear, quadratic and interaction terms of $x_i$ and $y_i$. The quadratic terms are necessary to incorporate the idea that a player is most likely to field a ball hit directly at him, and the interaction term captures the notion that it may be easier to make plays moving to one side (e.g. shortstops have better range moving to their left since they are moving towards first base).    Estimates of the coefficients $\hat{\btheta}_\ell$ are calculated from all balls in play %hit to the fielding position $\ell$
.    As an example, the surface of our fielding model for centerfielders is illustrated in Figure \ref{fig:cf}.

For ball in play $i$, we use the coordinates  ($x_i,y_i$) and the estimated coefficients $\hat{\btheta}_\ell$ for each fielding position $\ell$ to estimate the probability $\hat{p}_{i \ell}$ that fielder $\ell$ makes an out on ball in play $i$.   We normalize these probabilities across positions to estimate the responsibility $s_{i \ell}$ 
$$
	\hat{s}_{i \ell} = \frac{\hat{p}_{i \ell}}{\sum_{\ell} \hat{p}_{i \ell}},
$$
of the $\ell^{th}$ fielder on the $i^{th}$ play, which gives us the run value $\delta^f_i \cdot \hat{s}_{i \ell}$ for each fielder $\ell$.  Finally, we fit a regression model to adjust the fielding run values for the ballpark in which ball in play $i$ occurred, 
\begin{eqnarray}
\delta^f_i \cdot \hat{s}_{i \ell} = \bD_i \, \cdot \, \bphi  + \tau_{i\ell} 
\end{eqnarray}
where the covariate vector $\bD_i$ contains a set of indicator variables for the specific ballpark for plate appearance $i$. The coefficient vector $\bphi$ contains the effects of each ballpark which is estimated across all balls in play.  The estimated residuals of this model, 
\begin{eqnarray}
{\rm RAA}^{field}_{i\ell} = \hat{\tau}_{i\ell} = \delta^f_i \cdot \hat{s}_{i \ell} - \bD_i \,\cdot \, \widehat{\bphi}
\end{eqnarray}
represent the run value above average for fielder $\ell$ on ball in play $i$.  

\subsection{Pitching Run Values}\label{pitching}

In Section~\ref{defensive} above, we allocated run value $\delta^p_i$ to the pitcher for plate appearance $i$.   We need to adjust these run values to account for ballpark and platoon advantage, since both factors affect our expectation of pitching performance.   We fit the following regression model, 
\begin{eqnarray}
	\delta_{i}^p \,\, = \,\, \bB_i \,\cdot \, \bpsi  \,\, + \,\, \xi_i \,,  \label{reg9}
\end{eqnarray}
where the covariate vector $\bB_i$ contains a set of indicator variables for the specific ballpark for plate appearance $i$ and an indicator variable for whether or not the batter has a platoon advantage over the pitcher (same as in equation \ref{reg1}).  The coefficient vector $\bpsi$ contains the effects of each ballpark and the effect of a platoon advantage on the pitching run values.  We estimate the coefficients 
$\widehat{\bpsi}$ using the pitching run values for all plate appearances $i$.   The estimated residuals of this model, 
\begin{eqnarray}
{\rm RAA}^{pitch}_{i} = \hat{\xi}_{i} = \delta_{i}^p - \bB_i \,\cdot \, \widehat{\bpsi}
\end{eqnarray}
represent the run value above average for the pitcher on plate appearance $i$.  

\subsection{Tabulating Runs Above Average}
%GREG COMMENT: This is how I would do the summation notation, but it may be too much.  If you have a simpler way of doing this do it.  
As outlined in Sections~\ref{deltas}-\ref{pitching}, we can calculate  the run value for the hitter (${\rm RAA}^{hit}_{i}$), the run values for each baserunner (${\rm RAA}^{br}_{ij}$), the run values for each fielder (${\rm RAA}^{field}_{i\ell}$) and the run value for the pitcher (${\rm RAA}^{pitch}_{i}$) in each plate appearance $i$.  

The overall run value for a particular player $q$ is calculated by tabulating these run values across all plate appearances involving that player as a hitter, pitcher, baserunner or fielder,  
\begin{eqnarray*}
	RAA_q & = & \sum_{i} {\rm RAA}^{hit}_{i} \, \cdot \, {\rm I} \, ({\rm hitter} = q) \, +  \\ 
	& & \sum_j \sum_i {\rm RAA}^{br}_{ij} \, \cdot \, {\rm I} \, ({\rm runner} \, j = q) \, + \\ 
	&  & \sum_\ell \sum_i {\rm RAA}^{field}_{i \ell} \, \cdot \, {\rm I} \, ({\rm fielder} \, \ell = q) \, + \\
	&  & \sum_{i} {\rm RAA}^{pitch}_{i} \, \cdot \, {\rm I} \, ({\rm pitcher} = q)
\end{eqnarray*}

\begin{figure}
	\centering
	\includegraphics[width=0.9\textwidth]{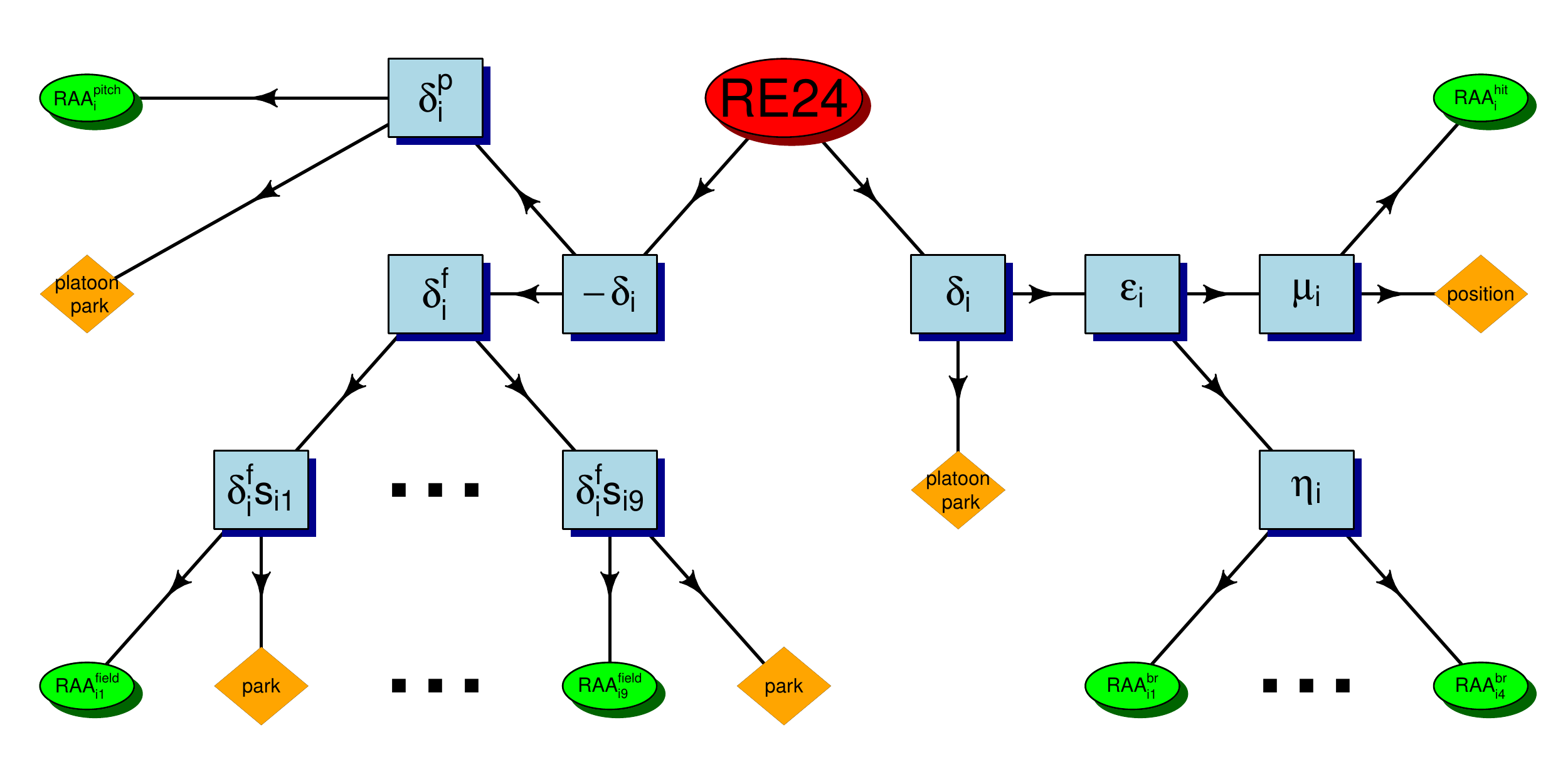}
	\caption{Schematic Diagram of $openWAR$:  The red ellipse labelled RE24 represents the estimated change in run value of a plate appearance.  This value is then split into many parts and atrributed to the appropriate source.  The diamonds represent fractions of RE24 that are not attributable to the player, whereas the ellipses on the outside correspond the four components of $openWAR$ (hitting, baserunning, pitching, and fielding) that are attributable to a player. }
	\label{fig:flowchart}
\end{figure}

We present a logical summary of our WAR calculation in Figure~\ref{fig:flowchart}.

\subsection{Replacement Level}
\label{sec:repl}

As noted in our introduction, it is desirable to calibrate our comprehensive measure of player performance relative to a baseline ``replacement level" player.  However, the definition of a replacement level player remains controversial. The procedure used by both the $fWAR$ and $rWAR$ implementations is to set replacement level ``at 1,000 WAR per 2,430 Major League games, which is the number of wins available in a 162 game season played by 30 teams. Or, an easier way to put it is that our new replacement level is now equal to a .294 winning percentage, which works out to 47.7 wins over a full season"~\citep{macaree}.  This definition is \emph{ad hoc}, with the primary motivation for the definition seems to be the use of a convenient round number. In contrast, we derive a natural definition for replacement level from first principles. 

The purpose of the replacement-level player is the need to \emph{replace} a full-time major league player.  There are only so many major league players, and all other players who participate in major league baseball are necessarily replacement players. Since there are 30 major league teams, each of which carries 25 players on its active roster during the season\footnote{In September, active rosters may expand to as many as 40 players, although in practice, few teams carry more than 30.}, there are exactly 750 active major league players on any given day.  We use this natural limitation to demarcate the set of major league players, and deem all others to be replacement-level players.  Since most teams carry 13 position players and 12 pitchers, we designate the $30 \cdot 13 = 390$ position players with the most plate appearances and the $30 \cdot 12 = 360$ pitchers with the most batters faced as major league players. We submit that this naturally-motivated definition of replacement level is preferable to the \emph{ad hoc} definition currently in use. 

We can associate a replacement-level \emph{shadow} with an actual player by multiplying the average performance across all replacement-level players by the number of events for that actual player. The WAR accumulated by each player's replacement-level shadow provides a meaningful baseline for comparison that is specific to that player. Using the convention that approximately 10 wins are equivalent to one win~\footnote{Justification for this convention is provided in our supplementary materials.}, our $openWAR$ value is computed as
$$
	WAR_q = \frac{RAA_q - RAA_q^{repl}}{10} \,,
$$
where $RAA_q^{repl}$ is the runs above average figure for player $q$'s replacement-level shadow. 

\section{Sources of Variability}
\label{sec:interval}

%\greg{Do we need notation for this?}
%In traditional computations of WAR, there are three main types of errors: measurement error, modeling error, and sampling error \greg{Is this what we want to call this piece?}. 

Existing implementations of WAR discuss uncertainty vaguely or not at all. 
We can delineate three sources of variability in the WAR values for each player in a given season: model estimation variability, situational variability, and player outcome variability.     Model estimation variability comes from the errors that are made in estimating the parameters of our models for batting, pitching, fielding and baserunning in Section~\ref{openwarmodel} as well as the expected runs model in Section~\ref{prelim}.    These models are trained on up to hundreds of thousands of observations and so this source of variability is small relative to the player outcome variability described below. 

Situational variability comes from the differences in game situations across occurrences of the same batting event.  For example, some home runs are hit when the bases are loaded whereas other home runs are hit when the bases are empty.  These two situations have very different run consequences despite the fact that they are driven by the same batting event (a home run).   In traditional $WAR$ calculations, a linear weights estimator is used for the batting component that assigns a run value to players based on aggregate batting statistics (such as $wOBA$) regardless of the game state.   In other words, all home runs are given the same value in traditional $WAR$ implementations, which introduces error into a player's $WAR$ value in the sense that an equal weighting of all home runs is a less accurate description of {\it what actually happened}.  (See \cite{wyers2013rw} for a discussion of quantifying situational error associated with $WARP$.)   In contrast, our $openWAR$ system is not subject to this type of error as we compute $WAR$ based on each plate appearance rather than using aggregate statistics, which is a key distinction from the three previous implementations of WAR ($fWAR$, $rWAR$ and $WARP$).  

Player outcome variability is the uncertainty inherent in the outcomes of all events involving a particular player for a particular season.  Imagine a particular player with a fixed ability, but repeat the same season for that player many times.  In each of these seasons, the events involving that player would have variation in their outcomes, which would aggregate to a different $WAR$ value for that particular player.   The random variation in individual events dominates the variability in a player's $WAR$ value, which is why our variance estimation targets this source of uncertainty.   

Specifically, we estimate player outcome variability using a resampling strategy.   In a particular season, we resample (with replacement) the $RAA$ values for \emph{individual plate appearances}, and re-aggregate them into a new $WAR$ value for each player.   A single resampling (a theoretical simulated season) will result in a second set of point estimates for the $WAR$ of each player for which the models have not changed but the number of different individual events (e.g. the number of home runs hit by a player) could have changed. By performing this resampling procedure many times, we quantify the outcome variability for each player while preserving any inherent correlation within the individual events.\footnote{We could use a similar resampling strategy to evaluate the model estimation variability as well, by re-fitting all of our $openWAR$ models on each resampled season.  However, the computational burden for re-fitting each model on each resampled season is quite high and hard to justify given the relatively small size of the model estimation variability compared to the player outcome variability.}.  Although we have discussed uncertainty specifically for $WAR$, we believe that the above delineation of variability sources is generalizable to most aggregate measures of player performance across sports.

\section{Results}

\begin{figure}
  \centering
	\includegraphics[width=\textwidth]{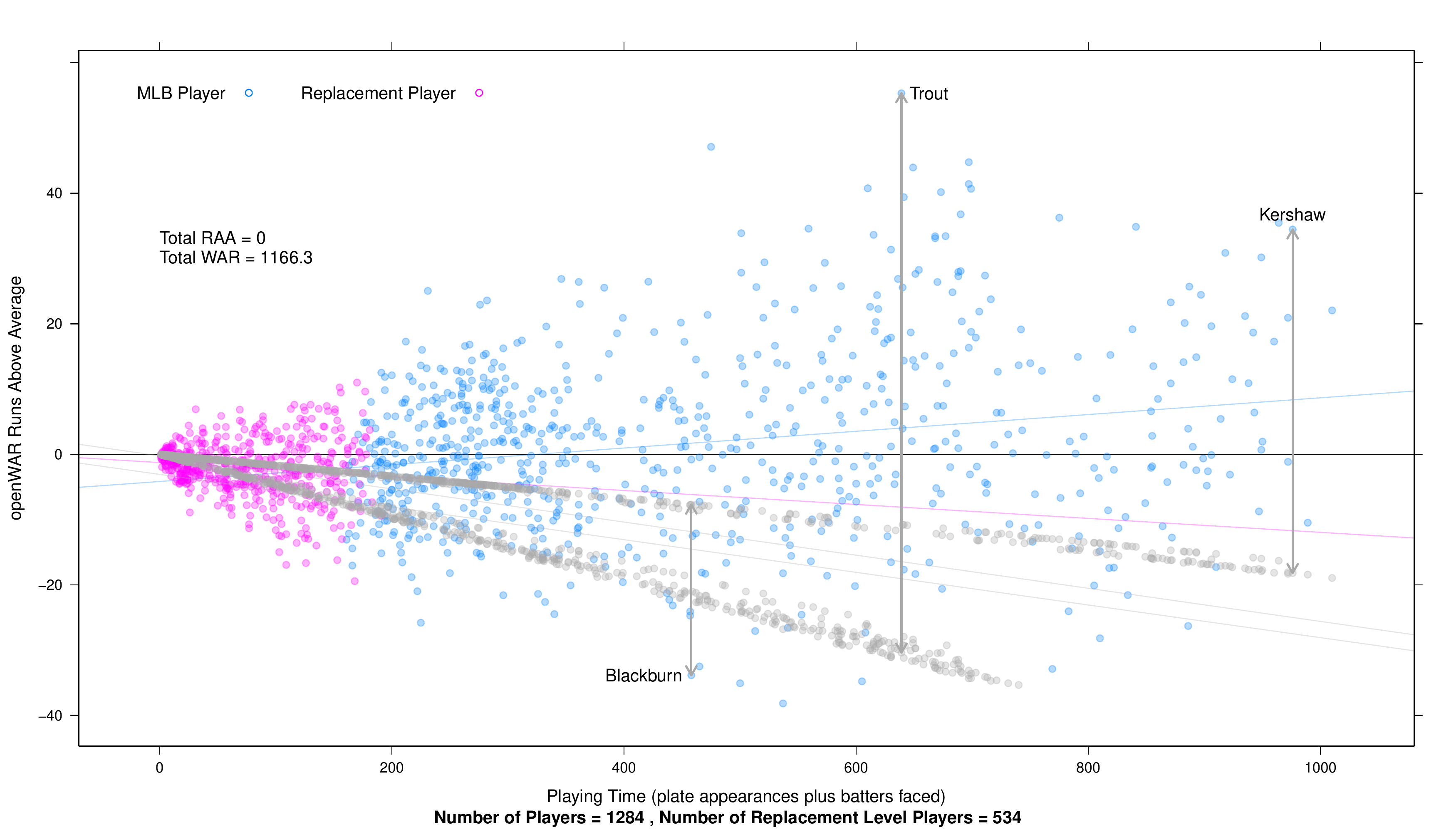}
  	\caption{$openWAR$ RAA values for the 2012 MLB season. Each blue dot is a major league player, while each pink dot is a replacement-level player. For each player, we also plot a gray dot that represents their replacement-level shadow with the same amount of playing time.  For three specific players, we show the vertical distance between their $RAA$ and the $RAA$ for their replacement-level shadow.  Playing time is calculated as ``plate appearances + batters faced" to provide an equivalent scale for both pitchers and batters: playing time for pitchers is the number of batters faced, whereas playing time for batters is the number of plate appearances.  For pitchers, we also add any plate appearances they had as a batter.}
  	\label{fig:openwar2012}
  \end{figure}

In the 2012 MLB season, 534 of the $N=1284$ players were designated as replacement-level. $openWAR$ was distributed approximately normally among these replacement-level players, with a mean of $0.01$ and a standard deviation of $0.41$. Conversely, the distribution of $openWAR$ across all players was skewed heavily to the right, reflecting the disproportionate amount of $openWAR$ accumulated by relatively few players. While the median $openWAR$ was close to zero ($0.36$), the mean was a bit larger ($0.91$). $openWAR$ values for all players fell between $-2.6$ and $8.6$ wins above replacement, giving a range of 10 wins between the best player (Mike Trout) and the worst (Nick Blackburn). In Figure \ref{fig:openwar2012}, we depict $openWAR$ values for 2012, illustrating each player's replacement-level shadow and differentiating the major league players from the replacement-level players. There are $2N$ dots in Figure \ref{fig:openwar2012}: $N$ non-gray dots representing the RAA values for actual players and $N$ gray dots representing the RAA values for the replacement-level shadows of those players.

\begin{figure}
  \centering
	\includegraphics[width=\textwidth]{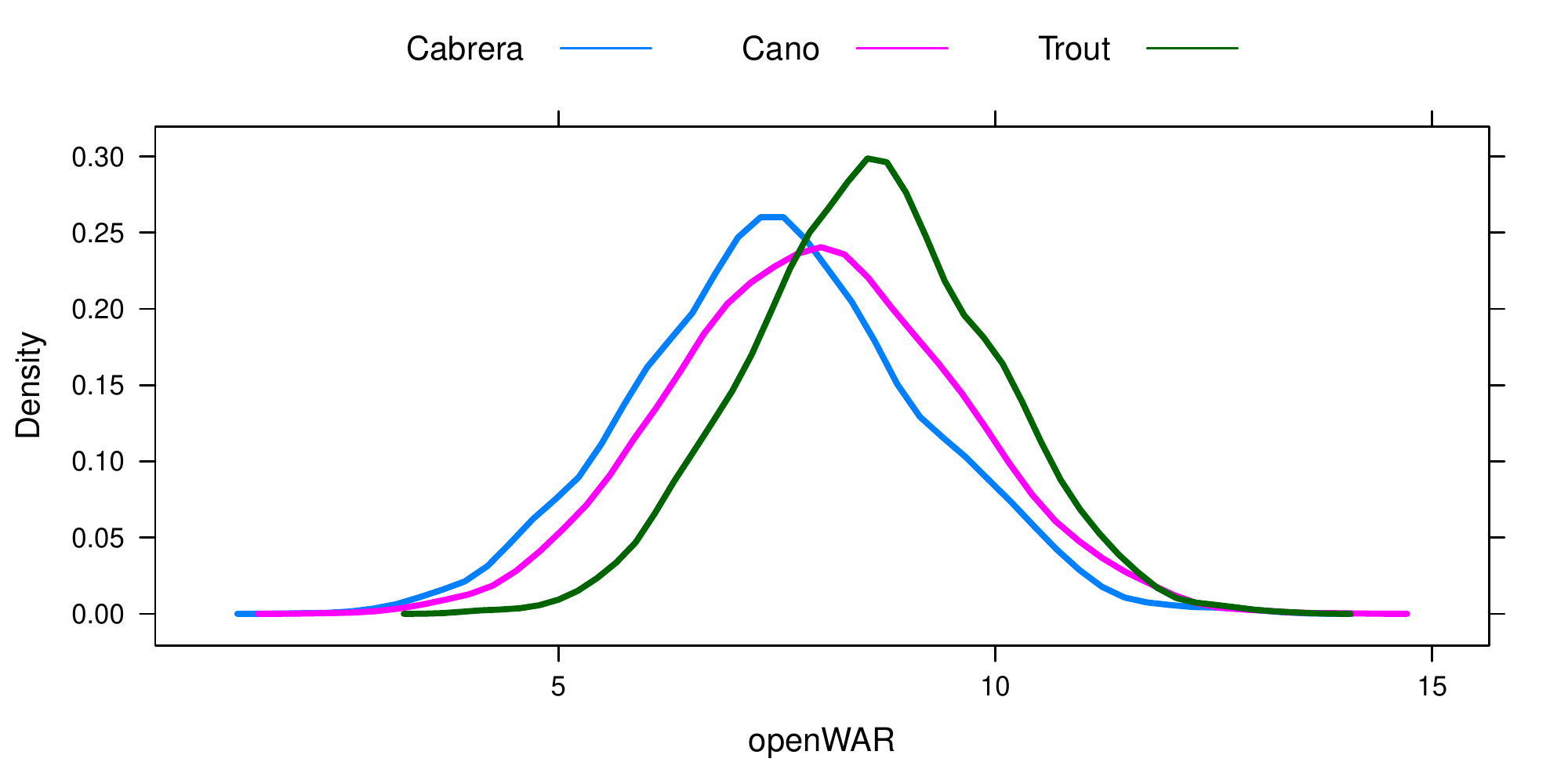}
  	\caption{$openWAR$ density estimates for Miguel Cabrera (blue), Robinson Cano (pink), and Mike Trout (green). Note that while Trout's density curve is further to the right, it is narrower than the others. }
  	\label{fig:int_est}
  \end{figure}

We note that the variability associated with player performance is not constant. Figure \ref{fig:int_est} shows density estimates for the distribution of $openWAR$ values under the resampling scheme described in Section~\ref{sec:interval} for three prominent players: Miguel Cabrera, Robinson Cano, and Mike Trout.  Trout's point estimate for WAR is higher than that of Cabrera or Cano, but the 95\% confidence interval for his true $openWAR$ is narrower, which suggests that Trout's performance is more consistent on a play-by-play basis than the others. Table \ref{tab:var} shows various quantiles of the distribution of $openWAR$ for the top 20 performers in 2012. 
  
% latex table generated in R 3.0.1 by xtable 1.7-1 package
% Fri Dec 20 17:21:39 2013
\begin{table}
\centering
\begin{tabular}{cccccccc}
  \hline
Name & q0 & q2.5 & q25 & q50 & q75 & q97.5 & q100 \\ 
  \hline
  Mike Trout & 3.52 & 5.81 & 7.58 & 8.53 & 9.48 & 11.27 & 13.91 \\ 
  Robinson Cano & 2.56 & 4.88 & 6.85 & 7.92 & 8.96 & 11.11 & 13.74 \\ 
  Chase Headley & 1.97 & 4.47 & 6.42 & 7.47 & 8.50 & 10.53 & 13.12 \\ 
  Miguel Cabrera & 1.99 & 4.31 & 6.43 & 7.46 & 8.49 & 10.49 & 12.84 \\ 
  Edwin Encarnacion & 2.67 & 4.54 & 6.36 & 7.32 & 8.29 & 10.29 & 13.17 \\ 
  Andrew McCutchen & 2.02 & 4.29 & 6.18 & 7.19 & 8.17 & 10.17 & 12.07 \\ 
  Joey Votto & 2.82 & 4.76 & 6.21 & 7.00 & 7.77 & 9.34 & 10.80 \\ 
  Prince Fielder & 2.57 & 4.16 & 5.98 & 6.96 & 7.90 & 9.89 & 12.18 \\ 
  Joe Mauer & 2.51 & 4.32 & 5.89 & 6.78 & 7.64 & 9.27 & 11.01 \\ 
  Buster Posey & 2.49 & 4.08 & 5.79 & 6.73 & 7.62 & 9.46 & 11.98 \\ 
  Aaron Hill & 1.29 & 3.72 & 5.64 & 6.62 & 7.59 & 9.52 & 12.67 \\ 
  Ryan Braun & 1.97 & 3.68 & 5.55 & 6.60 & 7.62 & 9.56 & 11.44 \\ 
  Ben Zobrist & 1.67 & 3.94 & 5.52 & 6.44 & 7.33 & 9.06 & 11.65 \\ 
  Josh Willingham & 0.83 & 3.33 & 5.25 & 6.29 & 7.27 & 9.44 & 11.75 \\ 
  Martin Prado & 1.59 & 3.69 & 5.27 & 6.16 & 7.05 & 8.65 & 10.97 \\ 
  Aramis Ramirez & 0.52 & 3.19 & 5.18 & 6.15 & 7.09 & 9.05 & 11.73 \\ 
  Elvis Andrus & 0.83 & 3.52 & 5.25 & 6.14 & 7.03 & 8.89 & 11.10 \\ 
  Matt Holliday & 0.54 & 3.22 & 5.07 & 6.09 & 7.09 & 9.02 & 11.20 \\ 
  Adrian Gonzalez & 1.40 & 3.04 & 4.91 & 5.93 & 6.96 & 8.84 & 10.89 \\ 
  David Wright & 1.20 & 3.22 & 4.95 & 5.88 & 6.81 & 8.64 & 10.63 \\ 
%  Beltre-A & 1.16 & 3.09 & 4.88 & 5.85 & 6.80 & 8.63 & 11.13 \\ 
%  Reyes-Js & 1.63 & 3.27 & 4.93 & 5.85 & 6.75 & 8.54 & 10.95 \\ 
%  Scutaro & 1.11 & 3.16 & 4.87 & 5.80 & 6.72 & 8.42 & 10.62 \\ 
%  Hamilton & 0.65 & 2.93 & 4.70 & 5.67 & 6.62 & 8.56 & 10.30 \\ 
%  Heyward & 0.67 & 3.05 & 4.73 & 5.66 & 6.54 & 8.29 & 10.85 \\ 
   \hline
\end{tabular}
\caption{Distribution of $openWAR$ for 2012. Quantiles reported are based on 3500 simulated seasons.} 
\label{tab:var}
\end{table}

\begin{figure}
  \centering
	\includegraphics[width=\textwidth]{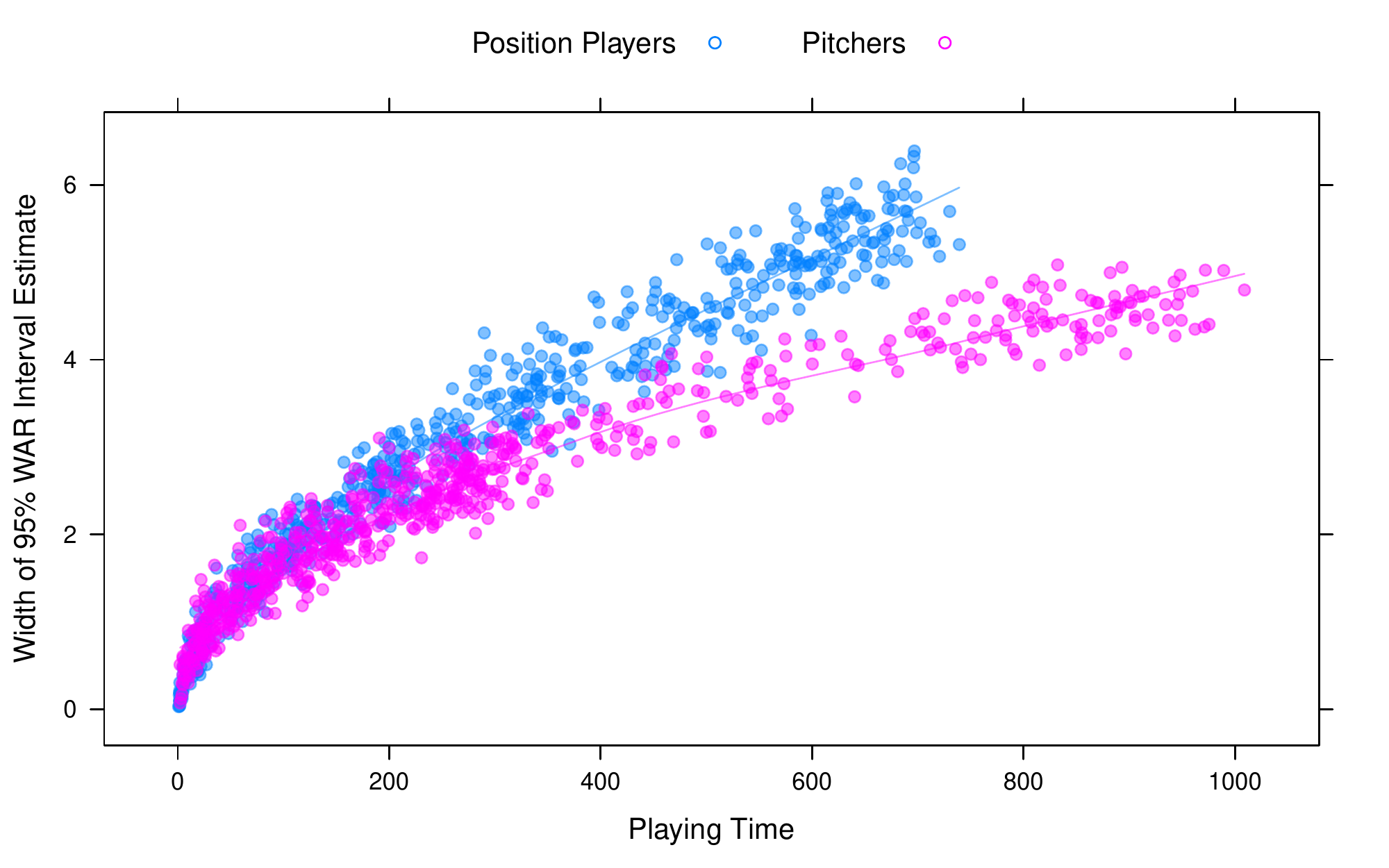}
  	\caption{Spread of 95\% Confidence Intervals for $openWAR$ based on resampling all plays that occurred in the 2012 season. Intervals are narrower for pitchers compared to position players and variation in the length of CIs exists among players. }
  	\label{fig:ci}
  \end{figure}

Figure \ref{fig:ci} depicts the width of 95\% confidence intervals for $openWAR$ based on resampling all plays that occurred in the 2012 season. As expected, the width of the confidence interval for a particular player widens as that player is exposed to more playing time. In general, the confidence intervals for pitchers tend to be smaller than those for position players with comparable playing time. This may suggest that pitchers perform more consistently across plate appearances, or merely reflect the fact that the replacement level for pitchers is higher (closer to 0) than it is for position players. 

As noted in the introduction, WAR was at the core of the debate about the 2012 American League MVP Award. Miguel Cabrera of the Detroit Tigers had become the first player since 1967 to win the Triple Crown, leading the AL in the conventional statistics of batting average, home runs, and runs batted in. However, sabermetricians advocated strongly for Mike Trout, a rookie centerfielder who excelled in all aspects of the game. While it was acknowledged on both sides that Cabrera was likely the better hitter, sabermetricians argued that Trout's superior skill at baserunning and fielding more than made up for Cabrera's relatively small edge in batting. In fact, for adherents of sabermetrics, the decision was clear -- point estimates showed Trout leading Cabrera by 3.2 $fWAR$ and 3.6 $rWAR$. 

\begin{figure}
  \centering
	\includegraphics[width=\textwidth]{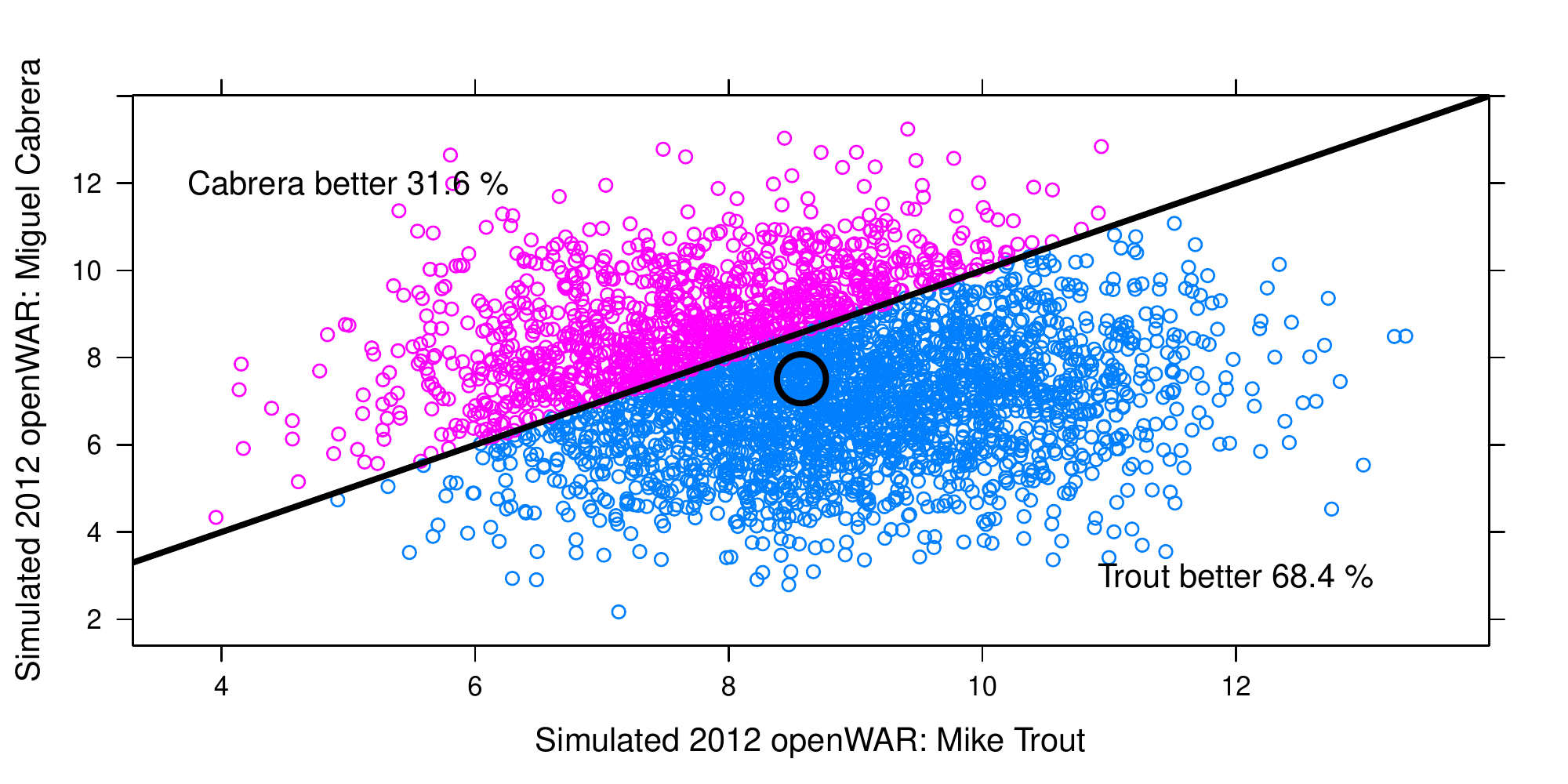}
  	\caption{Joint distribution of $openWAR$ for Mike Trout vs. Miguel Cabrera, 2012. We note that in about 68\% of 3500 simulated seasons, Trout produced a higher WAR than Cabrera.}
  	\label{fig:cvt}
  \end{figure}

Our $openWAR$ values provide a more sophisticated perspective on this debate. Trout's point estimate for $openWAR$ in 2012 is 1.05 wins larger than Cabrera's, but it is important to note that their interval estimates overlap considerably. In Figure \ref{fig:cvt}, the joint distribution of $openWAR$ values for Cabrera and Trout's 2012 seasons are plotted. In nearly 32\% of those simulated seasons, Cabrera's $openWAR$ was higher than Trout's. Thus, our results suggest that there is a high probability that Trout had a better season than Cabrera, but there is substantial uncertainty in their comparison.  This exercise illustrates two strengths of $openWAR$: 1) distinctions made through point estimates tend to accord with those made via the existing implementations (note that Cabrera was not even the second-best player in any implementation); and 2) the interval estimates provided by $openWAR$ allow for more nuanced conclusions to be drawn.  

Table \ref{tab:br} shows the top ten best and worst baserunners, according to $openWAR$ in 2012. We note many true positives (Mike Trout and Desmond Jennings are considered to be excellent baserunners, while Paul Konerko, David Ortiz and Adrian Gonzalez are plodding) with no eyebrow-raising surprises.

  \begin{table}
  	\centering
    \begin{tabular}{c|c||c|c|}
Best & $RAA$  & Worst & $RAA$\\
      \hline
Mike Trout & 14.79 & Paul Konerko & -9.28\\
Martin Prado & 9.03 & David Ortiz & -8.43\\
Desmond Jennings & 8.84 & Jamey Carroll & -8.27\\
Jarrod Dyson & 8.62 & Michael Young & -8.07\\
Evereth Cabrera &8.62 &Todd Helton & -7.08\\
Drew Stubbs & 7.86 &Prince Fielder &-6.82\\
Jason Heyward & 7.75 &Adrian Beltre & -6.34\\
Darwin Barney & 7.67 &Justin Morneau &-6.29\\
Torii Hunter & 7.64 &Adrian Gonzalez & -6.26\\
Dustin Ackley & 7.51& Howie Kendrick &-5.90\\
      \hline
    \end{tabular}
    \caption{2012 baserunning RAA Leaders}
    \label{tab:br}
  \end{table}
  
  Similarly, Table \ref{tab:fielding} shows the top ten best and worst fielders according to $openWAR$ in 2012. Here again we see some true positives (Brandon Crawford, Darwin Barney, and Adam Jones are reputedly excellent fielders) but also some head-scratchers (Prince Fielder is anecdotally considered a poor fielder). We also note that the magnitudes of the fielding numbers reported by $openWAR$ are smaller than those reported by UZR. This may be a result of the fact that $openWAR$ currently only measures some defensive skills, or it could reflect weaknesses in the unknown models underlying UZR, which merits further study.  

  \begin{table}
  	\centering
    \begin{tabular}{c|c||c|c|}
Best & $RAA$  & Worst & $RAA$\\
      \hline
Jason Heyward & 10.17 & Colby Rasmus & -10.44\\
Brandon Crawford & 9.91 & Jose Altuve & -9.03\\
Yunel Escober & 9.19 & Tyler Greene & -8.30\\
Ben Zobrist & 8.23 & Brian Dozier & -7.82\\
Darwin Barney &8.05 & Lucas Duda & -7.82\\
Prince Fielder & 7.66 & Shin-Soo Choo &-7.77\\
Adrian Gonzalez & 7.43 & Orlando Cespedes & -7.49\\
Alejandro De Aza & 7.13& Justin Smoak &-7.34\\
Adam Jones & 7.05 & Garrett Jones & -6.83\\
Craig Gentry & 6.72 & Rickie Weeks &-6.64\\
      \hline
    \end{tabular}
    \caption{2012 fielding RAA Leaders}
    \label{tab:fielding}
  \end{table}

\begin{figure}
  \centering
	\includegraphics[width=\textwidth]{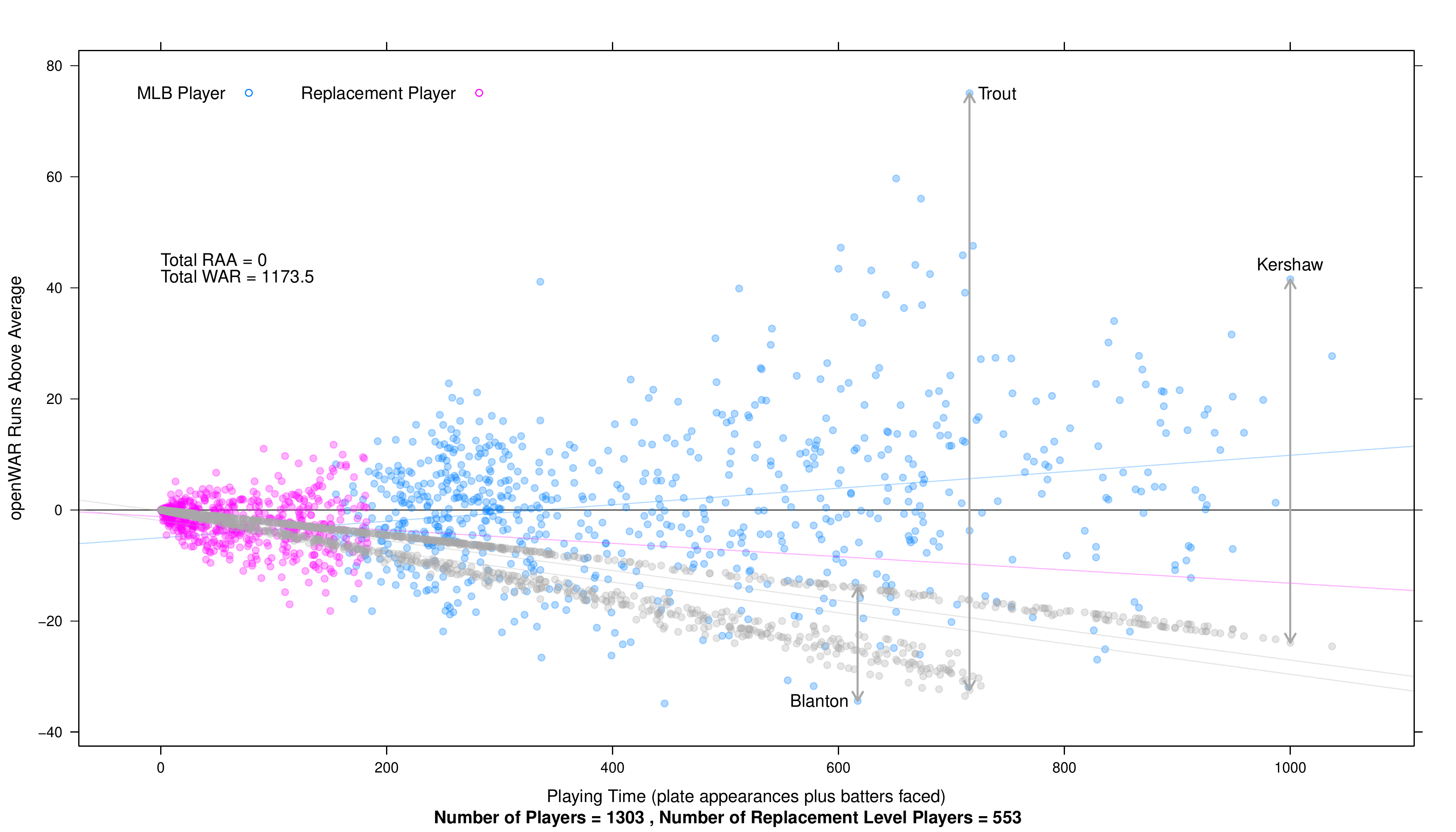}
  	\caption{$openWAR$ RAA values for the 2013 MLB season.  Each blue dot is a major league player, while each pink dot is a replacement-level player. For each player, we also plot a gray dot that represents their replacement-level shadow with the same amount of playing time.  Playing time is calculated just as in Figure~\ref{fig:openwar2012}.  Mike Trout and Clayton Kershaw were the best position player and pitcher, respectively, while Joe Blanton was the worst player.}
  	\label{fig:openwar2013}
  \end{figure}

  Results for $openWAR$ in the 2013 seasons were similar to those of 2012, with an observed range of $-2.0$ to $10.7$. Mike Trout was again the best player, and Clayton Kershaw was again the best pitcher ($6.5$ $openWAR$). Figure \ref{fig:openwar2013} shows the full $openWAR$ results for all players in 2013, and quantiles for simulated $openWAR$ for 2013 are presented in Table \ref{tab:var2013}.

% latex table generated in R 3.0.1 by xtable 1.7-1 package
% Mon Dec 23 09:26:38 2013
\begin{table}[ht]
\centering
\begin{tabular}{cccccccc}
  \hline
Name & q0 & q2.5 & q25 & q50 & q75 & q97.5 & q100 \\ 
  \hline
  Mike Trout & 5.48 & 7.60 & 9.58 & 10.60 & 11.60 & 13.58 & 15.39 \\ 
  Miguel Cabrera & 2.74 & 5.70 & 7.62 & 8.71 & 9.79 & 11.83 & 14.74 \\ 
  Chris Davis & 3.14 & 5.39 & 7.41 & 8.53 & 9.64 & 11.82 & 13.87 \\ 
  Matt Carpenter & 2.99 & 5.00 & 6.77 & 7.67 & 8.56 & 10.34 & 12.52 \\ 
  Paul Goldschmidt & 1.47 & 4.39 & 6.46 & 7.64 & 8.78 & 10.93 & 13.81 \\ 
  Josh Donaldson & 1.76 & 4.35 & 6.22 & 7.21 & 8.17 & 10.15 & 12.73 \\ 
  Matt Holliday & 2.07 & 4.34 & 6.16 & 7.07 & 7.97 & 9.89 & 12.58 \\ 
  Shin-Soo Choo & 2.63 & 4.33 & 6.14 & 7.06 & 7.98 & 9.89 & 12.02 \\ 
  Freddie Freeman & 1.13 & 4.28 & 6.03 & 7.05 & 8.04 & 9.93 & 11.59 \\ 
  Robinson Cano & 1.23 & 4.07 & 5.94 & 6.98 & 8.01 & 10.03 & 12.10 \\ 
  Andrew McCutchen & 1.75 & 4.00 & 5.74 & 6.71 & 7.70 & 9.53 & 11.80 \\ 
  David Ortiz & 1.61 & 3.87 & 5.60 & 6.63 & 7.63 & 9.59 & 12.08 \\ 
  Clayton Kershaw & 2.12 & 4.31 & 5.77 & 6.54 & 7.31 & 8.79 & 10.60 \\ 
  Carlos Santana & 2.35 & 3.89 & 5.54 & 6.42 & 7.30 & 8.89 & 11.04 \\ 
  Jason Kipnis & 1.68 & 3.62 & 5.35 & 6.29 & 7.23 & 9.04 & 11.23 \\ 
  Ian Kinsler & 1.17 & 3.33 & 5.06 & 5.92 & 6.79 & 8.47 & 10.76 \\ 
  Edwin Encarnacion & 1.03 & 3.12 & 4.94 & 5.91 & 6.84 & 8.90 & 11.71 \\ 
  Joey Votto & 1.43 & 3.31 & 4.96 & 5.91 & 6.82 & 8.63 & 10.44 \\ 
  Troy Tulowitzki & 1.15 & 3.33 & 5.02 & 5.88 & 6.75 & 8.47 & 10.04 \\ 
  Cliff Lee & 1.41 & 3.15 & 4.60 & 5.39 & 6.18 & 7.70 & 9.48 \\ 
%  Ellsbury & 0.26 & 2.66 & 4.47 & 5.39 & 6.33 & 8.10 & 10.22 \\ 
%  Ramirez-H & 1.26 & 3.26 & 4.59 & 5.35 & 6.09 & 7.55 & 9.67 \\ 
%  Bruce & 0.72 & 2.53 & 4.33 & 5.35 & 6.35 & 8.31 & 10.71 \\ 
%  Molina-Y & 1.53 & 3.03 & 4.49 & 5.34 & 6.11 & 7.79 & 9.67 \\ 
%  Gomez-C & 0.22 & 2.48 & 4.36 & 5.30 & 6.27 & 8.07 & 10.92 \\ 
   \hline
\end{tabular}
\caption{Distribution of $openWAR$ for 2013. Quantiles reported are based on 3500 simulated seasons.} 
\label{tab:var2013}
\end{table}

\subsection{Comparison to Previous WAR Implementations}

Our $openWAR$ point estimates are similar to existing implementations of WAR, though as noted above, we also provide uncertainty estimates. In Table \ref{tab:leaders}, we list the top 10 performance in $openWAR$ alongside those of $fWAR$. There is considerable (though not universal) agreement with respect to these players and the magnitudes of the WAR values are similar. Comparison to $rWAR$ yields similar results. 

%\greg{Correct intervals need to be filled in when we have them}
%  \begin{table}
%  	\centering
%    \begin{tabular}{c|c||c|c|c}
%      FanGraphs & $fWAR$ & [BJM] & $openWAR$ & 95\% CI \\
%      \hline
%Mike Trout & 10.0 & Mike Trout &  8.57 & $[5.81, 11.27]$ \\
%Robinson Cano & 7.7 & Robinson Cano &  7.91& $[4.88, 11.11]$ \\
%Buster Posey & 7.7 & Miguel Cabrera & 7.52 & $[4.31, 10.49]$ \\
%Ryan Braun & 7.6 & Chase Headley &  7.50 & $[4.47, 10.53]$ \\
%David Wright & 7.4 & Edwin Encarnacion & 7.28 & $[4.54, 10.29]$ \\
%Chase Headley & 7.2 & Andrew McCutchen & 7.24 & $[4.29, 10.17]$ \\
%Miguel Cabrera & 6.8 & Joey Votto & 6.96 & $[4.76, 9.34]$ \\
%Andrew McCutchen & 6.8 & Prince Fielder & 6.92  & $[4.16, 9.89]$ \\
%Jason Heyward & 6.4 & Joe Mauer & 6.73 & $[4.32, 9.27]$ \\
%Adrian Beltre & 6.3 & Buster Posey & 6.71 & $[4.08, 9.46]$ \\
%      \hline
%    \end{tabular}
%    \caption{2012  WAR Leaders, $fWAR$ (left) and $openWAR$ (right)}
%    \label{tab:leaders}
%  \end{table}
  
    \begin{table}
  	\centering
    \begin{tabular}{c|c||c|c}
      Name & $fWAR$ & Name & $openWAR$ \\
      \hline
Mike Trout & 10.0 & Mike Trout &  8.57 \\
Robinson Cano & 7.7 & Robinson Cano &  7.91 \\
Buster Posey & 7.7 & Miguel Cabrera & 7.52 \\
Ryan Braun & 7.6 & Chase Headley &  7.50 \\
David Wright & 7.4 & Edwin Encarnacion & 7.28 \\
Chase Headley & 7.2 & Andrew McCutchen & 7.24 \\
Miguel Cabrera & 6.8 & Joey Votto & 6.96 \\
Andrew McCutchen & 6.8 & Prince Fielder & 6.92 \\
Jason Heyward & 6.4 & Joe Mauer & 6.73 \\
Adrian Beltre & 6.3 & Buster Posey & 6.71 \\
      \hline
    \end{tabular}
    \caption{2012 WAR Leaders, $fWAR$ (left) and $openWAR$ (right)}
    \label{tab:leaders}
  \end{table}

We can examine the overall correlation between previous WAR implementations and our openWAR point estimates in Table \ref{tab:cor}.  $openWAR$ correlates highly with both $fWAR$ and $rWAR$, although not as highly as they correlate with each other. 

   \begin{table}
    \centering
    \begin{tabular}{cccc}
      & $rWAR$ & $fWAR$ & $openWAR$ \\
      \hline
      $rWAR$   & 1 & 0.918 & 0.881 \\
      $fWAR$   & 0.918 & 1 & 0.875 \\
      \hline
      \end{tabular}
     \caption{Correlation matrix between $openWAR$, $fWAR$, and $rWAR$.}
     \label{tab:cor}
   \end{table}
    
    We also examined the consistency of $openWAR$ from season-to-season by calculating the autocorrelation within players between their 2012 and 2013 seasons.  As seen in Table  \ref{tab:auto}, the within-player autocorrelation of our $openWAR$ values are similar to those of $fWAR$ and $rWAR$.
%Moreover, $openWAR$ is comparably \emph{reliable}. That is, it exhibits autocorrelation of the same order of magnitude as $fWAR$ and $rWAR$. These autocorrelations are shown in Table \ref{tab:auto}.
    
  \begin{table}
    \centering
    \begin{tabular}{cccc}
      & $rWAR$ & $fWAR$ & $openWAR$ \\
      \hline
      Autocorrelation & 0.522 & 0.596 & 0.571 \\
      \hline
    \end{tabular}
    \caption{Autocorrelation of WAR implementations. Each player's WAR in 2012 and 2013 was calculated, and the correlation between the matched pairs is shown.}
    \label{tab:auto}
  \end{table}

%In Table \ref{tab:leaders}, we present the leaders in Runs Above Average, according to the openWAR framework, alongside the corresponding figures from Fangraphs' $fWAR$. 
%
%  \begin{table}
%  	\centering
%    \begin{tabular}{c|c||c|c|c}
%      FanGraphs & $fRAA$ & [BJM] & $RAA$ & 95\% CI \\
%      \hline
%      Miguel Cabrera & 45.3 & Mike Trout & 54.5 & $[32.1, 77.2]$ \\
%      Mike Trout & 40.7 & Miguel Cabrera & 49.7 & $[24.8, 75.5]$ \\
%      Chris Davis & 36.6 & Chris Davis & 49.0 & $[25.2, 74.7]$ \\
%      David Wright & 33.6 & Jason Kipnis & 33.7 & $[13.5, 54.2]$ \\
%      Carlos Gomez & 33.7 & Troy Tulowitzki & 33.2 & $[14.4, 51.9]$ \\
%      Carlos Gonzalez & 30.8 & Paul Goldschmidt & 32.4 & $[9.9, 56.7]$ \\
%      Evan Longoria & 29.3 & David Ortiz & 32.2 & $[11.9, 54.7]$ \\
%      Manny Machado & 28.2 & Josh Donaldson & 32.2 & $[12.7, 53.6]$ \\
%      Josh Donaldson & 27.8 & Matt Carpenter & 31.0 & $[11.8, 50.8]$ \\
%      Matt Carpenter & 27.3 & Carlos Santana & 30.8 & $[11.7, 50.8]$ \\
%      \hline
%    \end{tabular}
%    \caption{2013 first-half WAR Leaders}
%    \label{tab:leaders}
%  \end{table}

\begin{figure}
  \centering
	\includegraphics[width=\textwidth]{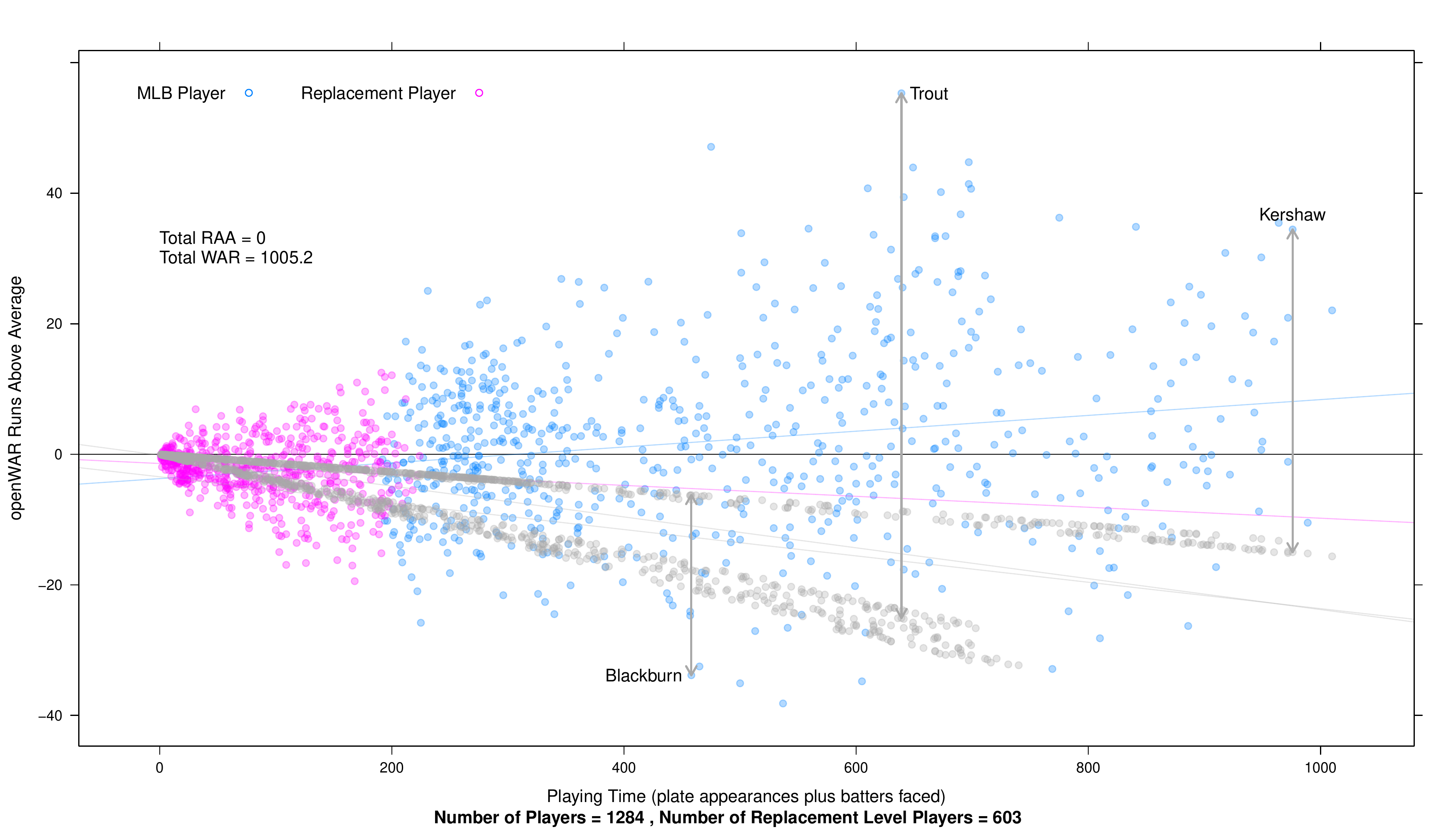}  							
	\caption{$openWAR$ RAA values for 2012, normalized so that the total WAR is about 1000. Compared to Figure \ref{fig:openwar2012}, here the definition of replacement-level is more inclusive.  Playing time is calculated just as in Figure~\ref{fig:openwar2012}.}
  	\label{fig:openwar2012n}
  \end{figure}

As illustrated in Figure \ref{fig:openwar2012}, the sum of all $RAA$ values in 2012 is exactly 0, and the sum of all $openWAR$ values is 1166. Whereas the former figure is guaranteed based on the way we have defined runs \emph{above average}, the latter is sensitive to changes in the definition of replacement-level. However, as noted in Section \ref{sec:repl}, replacement-level is defined in both $fWAR$ and $rWAR$ so that the sum of all WARs is 1000. In order to compare the magnitudes of $openWAR$ to $fWAR$ and $rWAR$ directly, we can generate more comparable values by increasing the number of replacement-level players. This in turn raises the performance of the replacement-level shadows, and lowers the amount of WAR in the system (see Figure \ref{fig:openwar2012n}). Given the \emph{ad hoc} nature of the previous definition of replacement-level, we prefer our definition. Moreover, the fact that the total unnormalized $openWAR$ was nearly identical in 2012 and 2013 (1166 and 1173), suggest that there may be some intrinsic meaning to this number. Additionally, the total RAA values for $rWAR$ did \emph{not} add up to 0 in either 2012 or 2013 -- a logical weakness in that system.

%In Figure \ref{fig:warplot_compare}, we illustrate how our uncertainty estimates inform a more nuanced understanding of WAR comparisons among players.   
%  
%  \begin{figure}
%  \centering
%	\includegraphics[width=\textwidth]{warplot_compare}
%  	\label{fig:warplot_compare}
%  \end{figure}

\section{Summary and Further Discussion}

The concept of Wins Above Replacement has been one of the great success stories in the long history of sabermetrics, and sports analytics in general. However, there are major limitations in previous methodology both in terms of calculating WAR, and in the public's understanding of what WAR values mean. Chiefly, the previous implementations of WAR are not reproducible and do not contain uncertainty estimates. This leads to the unpleasant situation where journalists are forced to take WAR estimates on faith, with no understanding of the accuracy (or construction) of those estimates. In this paper, we have addressed the issues of reproducibility and uncertainty estimation by providing a fully open source, statistical model for Wins Above Replacement based on our {\it conservation of runs} framework with uncertainty in our model-based WAR values estimated by resampling methods. 
  
There remain several limitations of $openWAR$ that offer the opportunity for further research.  The first limitation is data quality.  Although the fidelity of the MLBAM data is very high, it is not perfect.  There were instances in the data where players were listed with the wrong ID and for most balls in play, there was not a description of whether that ball was hit on the ground or in the air. Furthermore, there is no indication of how long each batted ball took to get to the specified location, making both the \emph{trajectory} and \emph{speed} of each batted ball unknown. 

The accounting of baserunner movement for non-batting events like stolen bases, caught stealings, wild pitches, and errors merits further work. All baserunner movement is captured, but it is modeled implicitly. Our approach only takes into account the actual baserunner movement but is indifferent to the various mechanisms by which a baserunner advanced. For example, a runner on first who steals second base and advances to third on a single is rewarded the same amount as a baserunner who advances to second on a wild pitch and then advances to third on a single. The same holds for a runner who simply advances directly from first to third on a single. 

The defensive models used in $openWAR$ are somewhat rudimentary and could certainly be improved with more resolute data. Since there is no record in the data of where each fielder was standing at the beginning of the play, there is no way to distinguish between fielder range versus fielder positioning. This drawback is also true in most current fielding measures, such as UZR and SAFE.   Some fielding measures such as UZR add additional components for throwing and the ability to turn a double play, which we hope to add to $openWAR$ in future work. 

Another interesting idea would be a conservation of wins framework for $openWAR$ rather than the conservation of runs. 
Rather than assigning the value of each plate appearance based on the change in expected runs, the value of a plate appearance could alternatively be based on the change in win probability from the beginning to the end of a plate appearance.  The $openWAR$ framework could then be altered to take changes in \emph{win probability} as inputs rather than the change in the expected run matrix.  One rationale for using win probability is that we may not wish to treat each run scored as contributing equally to a win.  For example, extra runs when a team is winning by a large margin are not as valuable as an extra run when the teams are tied.   

We suspect that using a framework based on the change in win probability will give similar results in terms of magnitude and ranking of players, since every day hitters will get plate appearances in many different game situations.  However, certain types of players (closers, relief specialists, pinch hitters, defensive replacements, pinch runners) may only make appearances in games in specific situations such that the runs that they create or prevent may be systematically more (or less) valuable than the approximately 10\% of a win that is assigned to each run now. It would be particularly interesting to look at relief pitchers as they are often only in a game because of the specific game situation (the game is close and near the end of the game), which would make the runs they prevent more valuable than most runs created over the course of the season.

%\section{Acknowledgements}
%We are grateful to Chadwick Baseball Bureau for use of their persons register.

%%%%%%%%%%%%%%%%%%%%%%%%%%%%%%%%%%%%%%%%%%%%%%%%%%%%%%%%%%%%%%%%%%%%%%%

\bibliographystyle{asa}
\bibliography{references}

\begin{thebibliography}{63}
\newcommand{\enquote}[1]{``#1''}
\expandafter\ifx\csname natexlab\endcsname\relax\def\natexlab#1{#1}\fi

\bibitem[{Acharya et~al.(2008)Acharya, Ahmed, D'Amour, Lu, Morris, Oglevee,
  Peterson, and Swift}]{acharya2008improving}
Acharya, R.~A., Ahmed, A.~J., D'Amour, A.~N., Lu, H., Morris, C.~N., Oglevee,
  B.~D., Peterson, A.~W., and Swift, R.~N. (2008), \enquote{Improving major
  league baseball park factor estimates,} \textit{Journal of Quantitative
  Analysis in Sports}, 4.

\bibitem[{Albert and Bennett(2003)}]{albert2003cbb}
Albert, J. and Bennett, J. (2003), \textit{{Curve Ball: Baseball, Statistics,
  and the Role of Chance in the Game}}, Copernicus Books.

\bibitem[{Appelman(2008)}]{re24}
Appelman, D. (2008), \enquote{Get to Know: RE24,}
  \url{http://www.fangraphs.com/blogs/get-to-know-re24/}.

\bibitem[{Appelman(2010)}]{appelmanuzr}
--- (2010), \enquote{UZR Updates!}
  \url{http://www.fangraphs.com/blogs/uzr-updates/}.

\bibitem[{Axisa(2013)}]{topps}
Axisa, M. (2013), \enquote{Topps will feature WAR on the back of their baseball
  cards soon,}
  \url{http://www.cbssports.com/mlb/eye-on-baseball/22903637/topps-will-feature-war-on-the-back-of-their-baseball-cards-soon}.

\bibitem[{{Baseball Prospectus Staff}(2013)}]{warp}
{Baseball Prospectus Staff} (2013), \enquote{WARP,}
  \url{http://www.baseballprospectus.com/glossary/index.php?search=warp}.

\bibitem[{Baumer and Matthews(2013)}]{openWAR}
Baumer, B. and Matthews, G.~J. (2013), \textit{openWAR: An Open Source System
  for Overall Player Performance in Major League Baseball},
  \url{http://github.com/beanumber/openWAR/}.

\bibitem[{Baumer and Zimbalist(2014)}]{baumer2013tsr}
Baumer, B. and Zimbalist, A. (2014), \textit{{The Sabermetric Revolution:
  Assessing the Growth of Analytics in Baseball}}, University of Pennsylvania
  Press.

\bibitem[{Bowman(2013)}]{mlbam}
Bowman, B. (2013), \enquote{GameDay,} \url{http://mlb.mlb.com/mlb/gameday}.

\bibitem[{Buckheit and Donoho(1995)}]{buckheit1995wavelab}
Buckheit, J.~B. and Donoho, D.~L. (1995), \enquote{Wavelab and reproducible
  research,} Tech. Rep. 474, Stanford University,
  \url{http://statistics.stanford.edu/~ckirby/techreports/NSF/EFS\%20NSF\%20474.pdf}.

\bibitem[{Bukiet et~al.(1997)Bukiet, Harold, and Palacios}]{bukiet1997mca}
Bukiet, B., Harold, E., and Palacios, J. (1997), \enquote{{A Markov Chain
  Approach to Baseball},} \textit{Operations Research}, 45, 14--23.

\bibitem[{Cameron(2008)}]{cameron2008win}
Cameron, D. (2008), \enquote{Win Values Explained: Part Five,}
  \url{http://www.fangraphs.com/blogs/win-values-explained-part-five/}.

\bibitem[{Claerbout(1994)}]{claerbout1994hypertext}
Claerbout, J. (1994), \enquote{Hypertext documents about reproducible
  research,} Tech. rep., Stanford University, \url{http://sepwww.stanford.edu}.

\bibitem[{Davenport and Woolner(1999)}]{davenport1999pythag}
Davenport, C. and Woolner, K. (1999), \enquote{Revisiting the Pythagorean
  Theorem,} \url{http://baseballprospectus.com/article.php?articleid=342}.

\bibitem[{{Fangraphs Staff}(2013)}]{drs}
{Fangraphs Staff} (2013), \enquote{DRS,}
  \url{http://www.fangraphs.com/library/defense/drs/}.

\bibitem[{Fast(2009)}]{fastdips}
Fast, M. (2009), \enquote{Confessions of a DIPS apostate,}
  \url{http://www.hardballtimes.com/confessions-of-a-dips-apostate/}.

\bibitem[{Forman(2010)}]{bbref}
Forman, S. (2010), \enquote{Player Wins Above Replacement,}
  \url{http://www.baseball-reference.com/blog/archives/6063}.

\bibitem[{Forman(2013{\natexlab{a}})}]{rwar}
--- (2013{\natexlab{a}}), \enquote{Position Player WAR Calculations and
  Details,}
  \url{http://www.baseball-reference.com/about/war_explained_position.shtml}.

\bibitem[{Forman(2013{\natexlab{b}})}]{compare}
--- (2013{\natexlab{b}}), \enquote{WAR Comparison Chart,}
  \url{http://www.baseball-reference.com/about/war_explained_comparison.shtml}.

\bibitem[{Freeze(1974)}]{freeze1974abb}
Freeze, R. (1974), \enquote{{An Analysis of Baseball Batting Order by Monte
  Carlo Simulation},} \textit{Operations Research}, 22, 728--735.

\bibitem[{Friendly(2013)}]{friendly}
Friendly, M. (2013), \textit{Lahman: Sean Lahman's Baseball Database}, r
  package version 2.0-3.

\bibitem[{Furtado(1999)}]{furtado1999xr}
Furtado, J. (1999), \enquote{{Introducing XR},}
  \url{http://www.baseballthinkfactory.org/btf/scholars/furtado/articles/IntroducingXR.htm}.

\bibitem[{Hamrahi(2013)}]{hamrahi}
Hamrahi, J. (2013), \enquote{Replacement Level and 10-Year Projections,}
  \url{http://www.baseballprospectus.com/article.php?articleid=19910}.

\bibitem[{Humphreys(2011)}]{humphreys2011wizardry}
Humphreys, M. (2011), \textit{Wizardry: Baseball's All-time Greatest Fielders
  Revealed}, Oxford University Press.

\bibitem[{Ioannidis(2013)}]{ioannidis2013}
Ioannidis, J.~P. (2013), \enquote{This I believe in genetics: discovery can be
  a nuisance, replication is science, implementation matters,}
  \textit{Frontiers in Genetics}, 4.

\bibitem[{Jacques(2007)}]{vorp}
Jacques, D. (2007), \enquote{Value Over Replacement Player,}
  \url{http://www.baseballprospectus.com/article.php?articleid=6231}.

\bibitem[{James(1986)}]{james1986bjh}
James, B. (1986), \textit{{The Bill James Historical Baseball Abstract}},
  Random House Inc.

\bibitem[{James and Henzler(2002)}]{james2002win}
James, B. and Henzler, J. (2002), \textit{Win shares}, STATS Pub.

\bibitem[{Jensen et~al.(2009)Jensen, Shirley, and Wyner}]{jensen2009bayesball}
Jensen, S.~T., Shirley, K.~E., and Wyner, A.~J. (2009), \enquote{Bayesball: A
  Bayesian hierarchical model for evaluating fielding in major league
  baseball,} \textit{The Annals of Applied Statistics}, 491--520.

\bibitem[{Johnson(2014)}]{johnson2014}
Johnson, G. (2014), \enquote{New truths that only one can see,} \textit{The New
  York Times},
  \url{http://www.nytimes.com/2014/01/21/science/new-truths-that-only-one-can-see.html}.

\bibitem[{Knuth(1984)}]{knuth1984literate}
Knuth, D.~E. (1984), \enquote{Literate programming,} \textit{The Computer
  Journal}, 27, 97--111.

\bibitem[{Kubatko et~al.(2007)Kubatko, Oliver, Pelton, and
  Rosenbaum}]{kubatko2007starting}
Kubatko, J., Oliver, D., Pelton, K., and Rosenbaum, D.~T. (2007), \enquote{A
  starting point for analyzing basketball statistics,} \textit{Journal of
  Quantitative Analysis in Sports}, 3, 1--22.

\bibitem[{Lahman(2013)}]{lahman}
Lahman, S. (2013), \enquote{Sean Lahman's Baseball Database,}
  \url{http://www.seanlahman.com/baseball-archive/statistics/}.

\bibitem[{Lewis(2003)}]{lewis2004maw}
Lewis, M. (2003), \textit{{Moneyball: The Art of Winning an Unfair Game}}, WW
  Norton \& Company.

\bibitem[{Lichtman(2010)}]{uzr}
Lichtman, M. (2010), \enquote{The Fangraphs UZR Primer,}
  \url{http://www.fangraphs.com/blogs/the-fangraphs-uzr-primer/}.

\bibitem[{Lichtman(2011)}]{ubr}
--- (2011), \enquote{Ultimate Base Running Primer,}
  \url{http://www.fangraphs.com/blogs/ultimate-base-running-primer/}.

\bibitem[{Lindsey(1959)}]{lindsey1959sdu}
Lindsey, G. (1959), \enquote{{Statistical Data Useful for the Operation of a
  Baseball Team},} \textit{Operations Research}, 7, 197--207.

\bibitem[{Lindsey(1961)}]{lindsey1961psd}
--- (1961), \enquote{{The Progress of the Score During a Baseball Game},}
  \textit{Journal of the American Statistical Association}, 56, 703--728.

\bibitem[{MacAree(2013)}]{macaree}
MacAree, G. (2013), \enquote{Replacement Level,}
  \url{http://www.fangraphs.com/library/misc/war/replacement-level/}.

\bibitem[{Macdonald(2011)}]{macdonald2011regression}
Macdonald, B. (2011), \enquote{A regression-based adjusted plus-minus statistic
  for NHL players,} \textit{Journal of Quantitative Analysis in Sports}, 7, 4.

\bibitem[{McCracken(2001)}]{mccracken01pd}
McCracken, V. (2001), \enquote{{Pitching and Defense: How Much Control Do
  Hurlers Have?}}
  \url{http://baseballprospectus.com/article.php?articleid=878}.

\bibitem[{Miller(2007)}]{miller2007derivation}
Miller, S.~J. (2007), \enquote{A derivation of the Pythagorean Won-Loss Formula
  in baseball,} \textit{Chance}, 20, 40--48.

\bibitem[{Naik(2011)}]{naik2011}
Naik, G. (2011), \enquote{Scientists' elusive goal: Reproducing study results,}
  \textit{The Wall Street Journal}, 258, A1,
  \url{http://online.wsj.com/news/articles/SB10001424052970203764804577059841672541590}.

\bibitem[{{Nature Editorial}(2013)}]{editorial2013}
{Nature Editorial} (2013), \enquote{Announcement: {R}educing our
  irreproducibility,} \textit{Nature}, 496,
  \url{http://www.nature.com/news/announcement-reducing-our-irreproducibility-1.12852}.

\bibitem[{Pankin(1978)}]{pankin1978eop}
Pankin, M. (1978), \enquote{{Evaluating Offensive Performance in Baseball},}
  \textit{Operations Research}, 26, 610--619.

\bibitem[{Rickert(2013)}]{rickert2013}
Rickert, J. (2013), \enquote{Nate Silver addresses assembled statisticians at
  this year's JSM,}
  \url{http://blog.revolutionanalytics.com/2013/08/nate-silver-jsm.html}.

\bibitem[{Rosenberg(2012)}]{rosenberg2012war}
Rosenberg, M. (2012), \enquote{The case for Miguel Cabrera over Mike Trout for
  AL MVP,}
  \url{http://sportsillustrated.cnn.com/2012/writers/michael_rosenberg/11/15/miguel-cabrara-mvp/index.html}.

\bibitem[{Schoenfield(2012)}]{schoenfield2012war}
Schoenfield, D. (2012), \enquote{What we talk about when we talk about WAR,}
  \url{http://espn.go.com/blog/sweetspot/post/_/id/27050/what-we-talk-about-when-we-talk-about-war}.

\bibitem[{Schwarz(2005)}]{schwarz2005ngb}
Schwarz, A. (2005), \textit{{The Numbers Game: Baseball's Lifelong Fascination
  With Statistics}}, Thomas Dunne Books.

\bibitem[{Slowinski(2010)}]{fangraphs}
Slowinski, S. (2010), \enquote{What is WAR?}
  \url{http://www.fangraphs.com/library/index.php/misc/war/}.

\bibitem[{Smith(2013)}]{retrosheet}
Smith, D. (2013), \enquote{Retrosheet,} \url{http://www.retrosheet.org/}.

\bibitem[{Sokol(2003)}]{sokol2003rhb}
Sokol, J. (2003), \enquote{{A Robust Heuristic for Batting Order Optimization
  Under Uncertainty},} \textit{Journal of Heuristics}, 9, 353--370.

\bibitem[{Studeman(2005)}]{studemanxfip}
Studeman, D. (2005), \enquote{I'm Batty for Baseball Stats,} \textit{The
  Hardball Times}, May 10, 2005,
  \url{http://www.fangraphs.com/library/pitching/xfip/}.

\bibitem[{Tango(2003)}]{tangofip}
Tango, T. (2003), \enquote{Defensive Responsibility Spectrum (DRS),}
  \url{http://www.tangotiger.net/drspectrum.html}.

\bibitem[{Tango(2008)}]{tango}
--- (2008), \enquote{How to calculate WAR,}
  \url{http://www.insidethebook.com/ee/index.php/site/article/how_to_calculate_war/}.

\bibitem[{Tango et~al.(2007)Tango, Lichtman, and Dolphin}]{tango2007bpp}
Tango, T., Lichtman, M., and Dolphin, A. (2007), \textit{{The Book: Playing the
  Percentages in Baseball}}, Potomac Books.

\bibitem[{{The Economist Editorial}(2013)}]{Economist2013}
{The Economist Editorial} (2013), \enquote{Trouble at the lab. (Cover story).}
  \url{http://www.economist.com/node/21588057/}.

\bibitem[{Thorn and Palmer(1984)}]{thorn1984hgb}
Thorn, J. and Palmer, P. (1984), \textit{{The hidden game of baseball: a
  revolutionary approach to baseball and its statistics}}, Doubleday, Garden
  City, NY.

\bibitem[{Tung(2010)}]{tung2010confidence}
Tung, D.~D. (2010), \enquote{Confidence Intervals for the Pythagorean Formula
  in Baseball,} \url{http://vixra.org/abs/1005.0020}.

\bibitem[{Wand(1994)}]{wand1994kernSmooth}
Wand, M. (1994), \enquote{Fast Computation of Multivariate Kernel Estimators,}
  \textit{Journal of Computational and Graphical Statistics}, 3, 433--445.

\bibitem[{Wyers(2013)}]{wyers2013rw}
Wyers, C. (2013), \enquote{Reworking WARP,}
  \url{http://www.baseballprospectus.com/article.php?articleid=21586}.

\bibitem[{Xie(2014)}]{xie2014knitr}
Xie, Y. (2014), \textit{{Dynamic Documents with R and knitr}}, Chapman \&
  Hall/CRC.

\bibitem[{Zimmer(2012)}]{zimmer2012}
Zimmer, C. (2012), \enquote{A Sharp Rise in Retractions Prompts Calls for
  Reform,} \textit{The New York Times},
  \url{http://www.nytimes.com/2012/04/17/science/rise-in-scientific-journal-retractions-prompts-calls-for-reform.html}.

\end{thebibliography}

\newpage

\phantom{xxxx}

\vspace{3cm}

\begin{center}
{\Large  {\bf Supplementary Materials for \\ 

\vspace{2cm}

``openWAR: An Open Source System for Evaluating Overall Player Performance in Major League Baseball"}}
\end{center}

\newpage

\appendix

\section{openWAR Package}

Code for the openWAR package is available for download on GitHub at \url{https://github.com/beanumber/openWAR}.

\section{Previous Implementations of WAR}
\label{sec:war_compare}

%There are many subtleties to each existing implementation of WAR, but 
The major components of each existing implementation of WAR are summarized in Table \ref{tab:compare}. The details of how each of these components is calculated are beyond what we can present here. Instructions for how to reproduce these numbers are illustrative, but not rigorous~\citep{fangraphs, bbref, rwar, compare, tango, uzr, wyers2013rw}.  At best, the authors may provide a step-by-step example calculation, but never specify a statistical model in formal notation, nor do they include code that would unambiguously reveal the algorithms used. The Baseball Info Solutions (BIS) data set, which is used to compute the fielding component of $rWAR$ and $fWAR$, is proprietary, and thus cannot be part of a reproducible piece of scholarship in which the results (as opposed to the models or algorithms) are the primary contribution. The high cost of obtaining this data (tens of thousands of dollars per year) prevents all but a few persons from verifying any results that stem from its use. The fielding metrics used by those two implementations, Defensive Run Saved ($DRS$) and Ultimate Zone Rating ($UZR$) are also proprietary. While extensive descriptions of each have been published~\citep{drs,uzr}, they too are illustrative---rather than specific---and none include source code. Occasionally, these organizations publish ``bug" fixes~\citep{hamrahi} or updates~\citep{appelmanuzr} that change previously published point estimates. Baseball Prospectus has announced plans to include more uncertainty and transparency in $WARP$~\citep{wyers2013rw}, but it is not known if this will include a release of source code\footnote{Incidentally, no further uncertainty updates to WARP have been published since Wyers left Baseball Prospectus to joined the Houston Astros in November 2013.}. 
	
	\begin{table}[ht!]
		\centering
		\begin{tabular}{c|c|c|c}
		 & $rWAR$ & $fWAR$ & $WARP$ \\
		 \hline
		 Data Source & BIS & BIS & Retrosheet \\
		 Batting 		& modified $wRAA$ & $wRAA$ & Linear Weights \\
		 Baserunning & Baserunning Runs & $BsR$ & Baserunning Runs Above Average \\
		 Fielding & $DRS$ & $UZR$ & Fielding Runs Above Average \\
		 Pitching & Runs Allowed & $FIP$ & Pitching Runs Above Average \\
		 \hline
		\end{tabular}
		\caption{Comparison of WAR implementations~\cite{compare}. The Baseball Info Solutions (BIS) data source is proprietary. Defensive Runs Saved ($DRS$) is a proprietary fielding metric developed by BIS. Ultimate Zone Rating ($UZR$) is a proprietary fielding metric developed by~\cite{uzr} and licesensed to Fangraphs.}
		\label{tab:compare}
	\end{table}

%	\begin{quotation}
%	Over the past four years, Mr. Zobrist has led baseball in WAR, ahead of stars like Albert Pujols, Ryan Braun and Robinson Cano.~\cite{eder2013war}
%	\end{quotation}	

\section{MLBAM data set}
\label{apx:data}

There are two main open sources of baseball data. \cite{lahman} maintains a database of seasonal data that has also been packaged for \R by \cite{friendly}. However, this data does not contain play-by-play information, making it insufficiently granular for WAR-type calculations, especially with respect to fielding. Retrosheet (\cite{retrosheet}) is an excellent source of free play-by-play data, but the batted ball locations are discrete, rather than continuous. That is, each batted ball is reported as falling into one of several dozen pre-defined polygonal zones. This level of detail is sufficient for some sophisticated defensive metrics, such as~\cite{humphreys2011wizardry}, but not others, such as UZR or SAFE~\citep{jensen2009bayesball}. Both of these data sources are updated periodically (usually at the end of the season). 

As noted in the paper, $openWAR$ uses data obtained from MLBAM. This data is not \emph{libre}, but it does reside on a publicly-available web server, making it \emph{gratis}. Furthermore, it is updated in real-time, and contains $(x,y)$-coordinates for each batted ball in every major league game. The \R package~\citep{openWAR} which has been developed simultaneously, will retrieve all data necessary to compute $openWAR$. %Because the data itself is not \emph{libre}, no data can be distributed with the package. However, t
The package contains simple \R functions that will enable any user with an Internet connection to download the data of their choice. 

The data available through this package is generally accurate. For example, summary statistics aggregated by team from all $184,739$ observations in 2012 are shown in Table \ref{tab:crosscheck} in the Appendix, next to the corresponding figures available through the Lahman database~\citep{lahman}. The agreement between the numbers presented in Table \ref{tab:crosscheck} is over 99.8\%\footnote{Specifically, the ratio of the Frobenius norm of the difference between the two sets and the Frobenius norm of the Lahman set is very small.}, indicating that the data collected and processed by $openWAR$ is of high fidelity.

%In Table \ref{tab:crosscheck}, we illustrate the accuracy of the MLBAM data set we collected by comparing aggregated results to those available in the Lahman database. The agreement between the two data sets is over 99.8\%. 

% latex table generated in R 3.0.1 by xtable 1.7-1 package
% Thu Aug 29 17:26:25 2013
\begin{table}[]
\centering
	\scalebox{0.73}{
\begin{tabular}{c|cccccccc|cccccccc}
  \hline
team & G & PA & AB & R & H & HR & BB & K & G & PA & AB & R & H & HR & BB & K \\ 
  \hline
ana & 162 & 6121 & 5537 & 766 & 1517 & 187 & 449 & 1112 & 162 & 6120 & 5536 & 767 & 1518 & 187 & 449 & 1113 \\ 
  ari & 162 & 6152 & 5466 & 734 & 1414 & 165 & 539 & 1266 & 162 & 6148 & 5462 & 734 & 1416 & 165 & 539 & 1266 \\ 
  atl & 162 & 6126 & 5427 & 699 & 1339 & 149 & 567 & 1289 & 162 & 6125 & 5425 & 700 & 1341 & 149 & 567 & 1289 \\ 
  bal & 162 & 6160 & 5562 & 712 & 1374 & 214 & 480 & 1315 & 162 & 6158 & 5560 & 712 & 1375 & 214 & 480 & 1315 \\ 
  bos & 162 & 6200 & 5636 & 737 & 1460 & 166 & 430 & 1204 & 162 & 6166 & 5604 & 734 & 1459 & 165 & 428 & 1197 \\ 
  cha & 162 & 6111 & 5518 & 748 & 1409 & 211 & 461 & 1202 & 162 & 6111 & 5518 & 748 & 1409 & 211 & 461 & 1203 \\ 
  chn & 162 & 5967 & 5411 & 613 & 1295 & 137 & 447 & 1235 & 162 & 5967 & 5411 & 613 & 1297 & 137 & 447 & 1235 \\ 
  cin & 162 & 6115 & 5477 & 669 & 1375 & 172 & 481 & 1266 & 162 & 6115 & 5477 & 669 & 1377 & 172 & 481 & 1266 \\ 
  cle & 162 & 6196 & 5526 & 667 & 1385 & 136 & 555 & 1087 & 162 & 6195 & 5525 & 667 & 1385 & 136 & 555 & 1087 \\ 
  col & 162 & 6183 & 5584 & 758 & 1525 & 166 & 450 & 1213 & 162 & 6176 & 5577 & 758 & 1526 & 166 & 450 & 1213 \\ 
  det & 162 & 6119 & 5477 & 726 & 1465 & 163 & 511 & 1103 & 162 & 6119 & 5476 & 726 & 1467 & 163 & 511 & 1103 \\ 
  hou & 162 & 6014 & 5409 & 583 & 1276 & 146 & 463 & 1364 & 162 & 6012 & 5407 & 583 & 1276 & 146 & 463 & 1365 \\ 
  kca & 162 & 6151 & 5638 & 676 & 1492 & 131 & 404 & 1032 & 162 & 6149 & 5636 & 676 & 1492 & 131 & 404 & 1032 \\ 
  lan & 162 & 6091 & 5438 & 637 & 1367 & 116 & 481 & 1156 & 162 & 6091 & 5438 & 637 & 1369 & 116 & 481 & 1156 \\ 
  mia & 162 & 6059 & 5440 & 610 & 1329 & 137 & 484 & 1228 & 162 & 6056 & 5437 & 609 & 1327 & 137 & 484 & 1228 \\ 
  mil & 162 & 6226 & 5559 & 776 & 1443 & 202 & 466 & 1240 & 162 & 6224 & 5557 & 776 & 1442 & 202 & 466 & 1240 \\ 
  min & 162 & 6209 & 5562 & 701 & 1446 & 131 & 505 & 1069 & 162 & 6209 & 5562 & 701 & 1448 & 131 & 505 & 1069 \\ 
  nya & 162 & 6231 & 5524 & 803 & 1462 & 245 & 565 & 1175 & 162 & 6231 & 5524 & 804 & 1462 & 245 & 565 & 1176 \\ 
  nyn & 162 & 6091 & 5454 & 650 & 1356 & 139 & 503 & 1250 & 162 & 6089 & 5450 & 650 & 1357 & 139 & 503 & 1250 \\ 
  oak & 162 & 6187 & 5532 & 714 & 1317 & 195 & 550 & 1386 & 162 & 6183 & 5527 & 713 & 1315 & 195 & 550 & 1387 \\ 
  phi & 162 & 6174 & 5546 & 684 & 1413 & 158 & 454 & 1094 & 162 & 6172 & 5544 & 684 & 1414 & 158 & 454 & 1094 \\ 
  pit & 162 & 6014 & 5412 & 651 & 1311 & 170 & 444 & 1354 & 162 & 6014 & 5412 & 651 & 1313 & 170 & 444 & 1354 \\ 
  sdn & 162 & 6112 & 5425 & 651 & 1336 & 121 & 539 & 1237 & 162 & 6112 & 5422 & 651 & 1339 & 121 & 539 & 1238 \\ 
  sea & 162 & 6061 & 5499 & 621 & 1285 & 149 & 466 & 1259 & 162 & 6057 & 5494 & 619 & 1285 & 149 & 466 & 1259 \\ 
  sfn & 162 & 6200 & 5559 & 718 & 1492 & 103 & 483 & 1097 & 162 & 6200 & 5558 & 718 & 1495 & 103 & 483 & 1097 \\ 
  sln & 162 & 6326 & 5624 & 765 & 1524 & 159 & 533 & 1192 & 162 & 6326 & 5622 & 765 & 1526 & 159 & 533 & 1192 \\ 
  tba & 162 & 6106 & 5401 & 697 & 1289 & 175 & 571 & 1324 & 162 & 6103 & 5398 & 697 & 1293 & 175 & 571 & 1323 \\ 
  tex & 162 & 6216 & 5592 & 808 & 1523 & 200 & 478 & 1103 & 162 & 6214 & 5590 & 808 & 1526 & 200 & 478 & 1103 \\ 
  tor & 162 & 6137 & 5525 & 723 & 1353 & 200 & 478 & 1255 & 162 & 6093 & 5487 & 716 & 1346 & 198 & 473 & 1251 \\ 
  was & 162 & 6221 & 5615 & 729 & 1467 & 194 & 479 & 1325 & 162 & 6221 & 5615 & 731 & 1468 & 194 & 479 & 1325 \\ 
   \hline
\end{tabular}
}
\caption{Cross-check between MLBAM data collected by openWAR (left) and Lahman data (right), 2012. These data are aggregated by team from 187,739 observations. } 
\label{tab:crosscheck}
\end{table}

In Table \ref{tab:events}, we list the 31 event types in the MLBAM data set, along with their frequencies of occurrence in 2012. 

% latex table generated in R 3.0.1 by xtable 1.7-1 package
% Wed Sep 11 11:55:58 2013
\begin{table}[]
\centering
\begin{tabular}{lcc}
  \hline
Event Type & $N$ & Frequency \\ 
  \hline
Strikeout & 36286 & 0.196 \\ 
  Groundout & 35266 & 0.191 \\ 
  Single & 27954 & 0.151 \\ 
  Flyout & 24890 & 0.135 \\ 
  Walk & 13660 & 0.074 \\ 
  Pop Out & 9072 & 0.049 \\ 
  Double & 8221 & 0.045 \\ 
  Lineout & 6666 & 0.036 \\ 
  Home Run & 4937 & 0.027 \\ 
  Forceout & 3984 & 0.022 \\ 
  Grounded Into DP & 3613 & 0.020 \\ 
  Field Error & 1705 & 0.009 \\ 
  Hit By Pitch & 1494 & 0.008 \\ 
  Sac Bunt & 1478 & 0.008 \\ 
  Sac Fly & 1213 & 0.007 \\ 
  Intent Walk & 1056 & 0.006 \\ 
  Triple &  927 & 0.005 \\ 
  Double Play &  494 & 0.003 \\ 
  Runner Out &  463 & 0.003 \\ 
  Bunt Groundout &  410 & 0.002 \\ 
  Fielders Choice Out &  352 & 0.002 \\ 
  Bunt Pop Out &  209 & 0.001 \\ 
  Strikeout - DP &  146 & 0.001 \\ 
  Fielders Choice &  114 & 0.001 \\ 
  Fan interference &   46 & 0.000 \\ 
  Batter Interference &   35 & 0.000 \\ 
  Catcher Interference &   23 & 0.000 \\ 
  Sac Fly DP &   11 & 0.000 \\ 
  null &    5 & 0.000 \\ 
  Bunt Lineout &    4 & 0.000 \\ 
  Triple Play &    3 & 0.000 \\ 
  Sacrifice Bunt DP &    2 & 0.000 \\ 
   \hline
\end{tabular}
\caption{Frequency of Events in MLBAM data set (2012)} 
\label{tab:events}
\end{table}

Nevertheless, there are some significant limitations to this data set~\citep{fastdips}. It is important to note that these data are collected for the purposes of entertainment (e.g. feeding the MLBAM web application) and not for the purposes of data analysis.

\section{Converting Runs to Wins}
\label{apx:pythag}

As changes in the run expectancy matrix are measured in \emph{runs}, but the units of WAR are \emph{wins}, it is necessary to convert runs to wins. A common convention used by all providers of WAR is that 10 runs is equivalent to 1 win~\citep{cameron2008win}. This value can thought of as a slope in the relationship between runs and wins at a point representing the average team. More specifically, this value can be derived as the partial derivative of Pythagorean Win Expectation evaluated at a specific point. 

Consider the general form of James' formula for \emph{expected winning percentage}, which is derivable if run scoring follows independent Weibull distributions~\citep{miller2007derivation}. That is, with $p$ equal to a parameter (originally 2), then 
$$
	WPct_p(RS, RA) = \frac{1}{1 + \left( \frac{RA}{RS} \right)^p}
$$
where $RS$ and $RA$ are the runs scored and allowed by a team, respectively. The gradient of this function is
$$
	\nabla WPct_p(RS, RA) = \left\langle \frac{\partial WPct_p}{\partial RS} , \frac{\partial WPct_p}{\partial RA} \right\rangle = \frac{p \cdot (RA/RS)^p}{(1 + (RA/RS)^p)^2} \cdot \left\langle \frac{1}{RS}, -\frac{1}{RA} \right\rangle \,.
$$
Thus, if $r=RS=RA$ (as it will be for an average team), then this becomes:
$$
	\nabla WPct_p (r,r) = \frac{p}{4} \left\langle \frac{1}{r}, -\frac{1}{r} \right\rangle = \frac{p}{4r} \cdot \langle 1, -1 \rangle \,.
$$
The gradient points in the direction of scoring more runs and allowing fewer, and from the magnitude we recover that the number of runs associated with one win over a 162 game season is:
$$
	\text{Runs per Win}_p(r) = \left( \frac{p}{4r} \right)^{-1} / 162 = \frac{2r}{81p}.
$$	
The optimal choice of the parameter $p$ may depend on the run-scoring environment. While James originally chose $p=2$ for convenience, better fits for Major League Baseball have been obtained using $p=1.83$~\citep{davenport1999pythag} and $p=1.86$~\citep{tung2010confidence}. The average number of runs scored per 162 games has been approximately $r=714$ since 1901, and $r=761$ since the league expanded to 30 teams in 1998. Reasonable choices for $p$ and $r$ will yield conversion factors in the neighborhood of 10.

\end{document}